\documentclass[twocolumn,times]{aastex62}

\shorttitle{Uncertainties in Atomic Data}
\shortauthors{Yu, Del Zanna, Stenning, et al.}

\usepackage{color}
\usepackage{amsmath}
\usepackage{enumerate}
\usepackage[figuresright]{rotating}

\usepackage[normalem]{ulem}

\newcommand{\calN}{\mathcal{N}}

\newcommand{\Data}{\mathcal{D}}
\newcommand{\data}{D}
\newcommand{\datak}{D_k}
\newcommand{\paramj}{\Theta}
\newcommand{\param}{\theta}
\newcommand{\paramk}{\theta_k}
\newcommand{\Ne}{n_e}

\newcommand{\ds}{ds}
\newcommand{\Nek}{n_{ek}}
\newcommand{\Tek}{T_{ek}}
\newcommand{\dsk}{ds_k}
\newcommand{\m}{m}
\newcommand{\rvm}{m}

\newcommand{\emisk}{\epsilon_{\lambda}^{(\m)}(\Nek, \Tek)}
\newcommand{\emiskj}{\epsilon_{\lambda_j}^{(\m)}(\Nek, \Tek)}
\newcommand{\emisNOm}{\epsilon_{\lambda}(\Nek, \Tek)}

\makeatletter
\newcommand{\distas}[1]{\mathbin{\overset{#1}{\kern\z@\sim}}}%
\newsavebox{\mybox}\newsavebox{\mysim}
\newcommand{\distras}[1]{%
  \savebox{\mybox}{\hbox{\kern3pt$\scriptstyle#1$\kern3pt}}%
  \savebox{\mysim}{\hbox{$\sim$}}%
  \mathbin{\overset{#1}{\kern\z@\resizebox{\wd\mybox}{\ht\mysim}{$\sim$}}}%
}
\makeatother

\ifpdf
\DeclareGraphicsExtensions{.pdf,.png,.jpg}
\fi

\begin{document}

%% ------------------------------------------------------------------------------------------
%% --- TITLE PAGE ---------------------------------------------------------------------------
%% ------------------------------------------------------------------------------------------

\title{Incorporating Uncertainties in Atomic Data Into the Analysis of Solar and Stellar
  Observations: A Case Study in \ion{Fe}{13}}

\author{Xixi Yu}
\affil{Statistics Section, Department of Mathematics, Imperial College London, London SW7 2AZ, UK}

\author{Giulio Del Zanna}
\affil{DAMTP, Centre for Mathematical Sciences, University of Cambridge, Wilberforce Road, Cambridge
  CB3 0WA, UK}

\author{David C.\ Stenning}
\affil{Statistics Section, Department of Mathematics, Imperial College London, London SW7 2AZ, UK}

\author{Jessi Cisewski-Kehe}
\affil{Department of Statistics and Data Science, Yale University, New Haven, CT  06511, USA}

\author{Vinay L.\  Kashyap}
\affil{Harvard-Smithsonian Center for Astrophysics, 60 Garden Street, Cambridge, MA 02138, USA}

\author{Nathan Stein}
\affil{Current Address: Spotify, 45 W 18th St, New York, NY 10011, USA}

\author{David~A.~van~Dyk}
\affil{Statistics Section, Department of Mathematics, Imperial College London, London SW7 2AZ, UK}

\author{Harry P. Warren}
\affil{Space Science Division, Naval Research Laboratory, Washington, DC 20375, USA}

\author{Mark A. Weber}
\affil{Harvard-Smithsonian Center for Astrophysics, 60 Garden Street, Cambridge, MA 02138, USA}

%% ------------------------------------------------------------------------------------------
%% --- ABSTRACT -----------------------------------------------------------------------------
%% ------------------------------------------------------------------------------------------

\begin{abstract}
Information about the physical properties of astrophysical objects cannot be measured directly but
is inferred by interpreting spectroscopic observations in the context of atomic physics
calculations. Ratios of emission lines, for example, can be used to infer the electron density of
the emitting plasma. Similarly, the relative intensities of emission lines formed over a wide range
of temperatures yield information on the temperature structure. A critical component of this
analysis is understanding how uncertainties in the underlying atomic physics propagates to the
uncertainties in the inferred plasma parameters. At present, however, atomic physics databases do
not include uncertainties on the atomic parameters and there is no established methodology for
using them even if they did. In this paper we develop simple models for the uncertainties in the
collision strengths and decay rates for \ion{Fe}{13} and apply them to the interpretation of
density sensitive lines observed with the EUV Imagining spectrometer (EIS) on \textit{Hinode}.  We
incorporate these uncertainties in a Bayesian framework.  We consider both a \textit{pragmatic
Bayesian} method where the atomic physics information is unaffected by the observed data, and
a \textit{fully Bayesian} method where the data can be used to probe the physics. The former
generally increases the uncertainty in the inferred density by about a factor of 5 compared with
models that incorporate only statistical uncertainties.  The latter reduces the uncertainties on
the inferred densities, but identifies areas of possible systematic problems with either the atomic
physics or the observed intensities.
\end{abstract}


%% ------------------------------------------------------------------------------------------
%% --- BODY ---------------------------------------------------------------------------------
%% ------------------------------------------------------------------------------------------

\section{Introduction}

Spectral observations of solar and stellar coronae, mostly taken in the X-ray, extreme ultraviolet
(EUV), and ultraviolet (UV) part of the spectrum, are regularly combined with atomic data to infer
fundamental plasma parameters such as electron temperatures and densities. This information is
essential for constraining models of coronal heating. In general in astrophysics, reliable atomic
data are essential for interpreting and modeling x-ray observations, as, for example, discussed in
\cite{2007RvMP...79...79K}. One key aspect about the modelling is the accuracy of the atomic data,
an area which has recently received some attention, see
e.g. \cite{2012IAUS..283..139L,2013ApJ...770...15B,2013AIPC.1545..242L,2016JPhD...49J3002C}.  In
these studies, guidelines to estimate uncertainties as a routine part of the computations of data
have been provided, or some preliminary analysis based on comparisons between different
calculations.

Over the years, the accuracy of spectral observations and of the atomic calculations has progressed
hand in hand. Current space instruments now provide measurements with accuracy of 20\% or better.
With the earliest observations and atomic data, discrepancies between measured and predicted line
intensities of factors of two or more were common. In the past years, thanks to large-scale atomic
structure and scattering calculations \citep[see, e.g. the
  reviews][]{2017Atoms...5...16J,2016JPhB...49i4001B}, the atomic data have improved
significantly. In a series of papers, starting from \cite{delzanna_etal:04_fe_10}, one of us (GDZ)
has benchmarked the available atomic data for several of the main ions against all available
experimental data, from laboratory sources to astrophysical spectra, and found generally good
agreement (within 10--20\%) between observations and theoretical calculations. Several reasons for
the remaining discrepancies have been identified.  Sometimes lines were blended, sometimes their
identifications were incorrect.  In some other cases the radiometric calibration of an instrument
was at fault, or the atomic calculations were incorrect.

\begin{figure*}[t!]
  \centerline{
    \includegraphics[width=\linewidth]{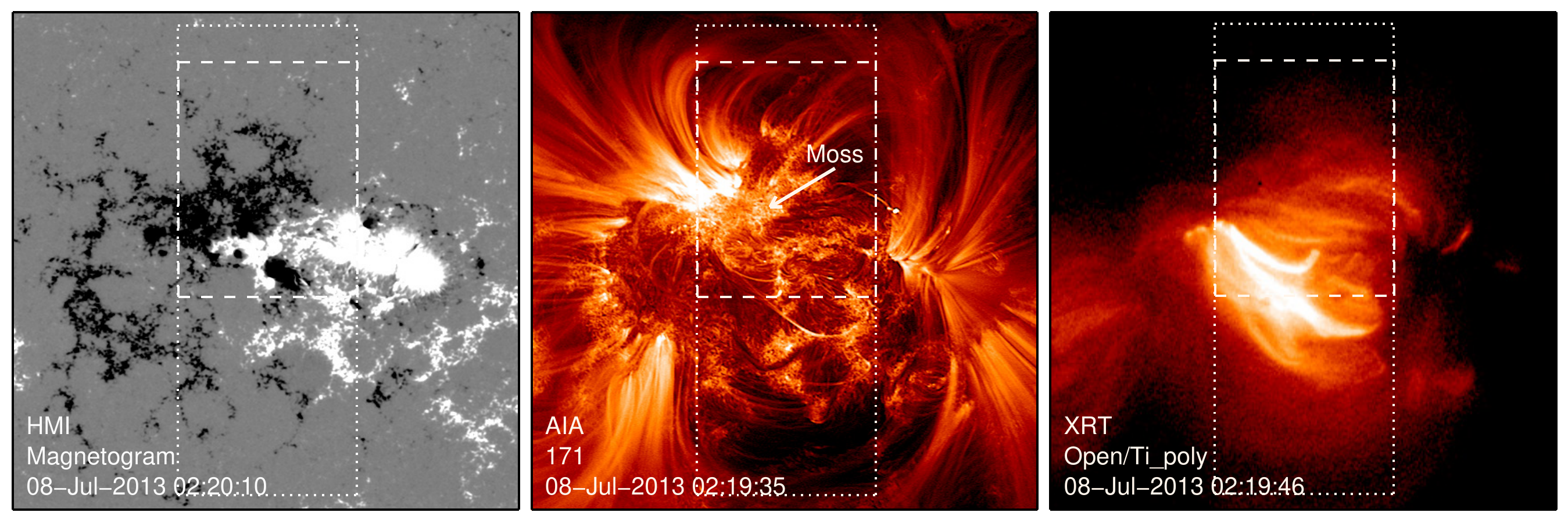}}
  \vspace{-1.5mm}
  \centerline{
    \includegraphics[width=\linewidth]{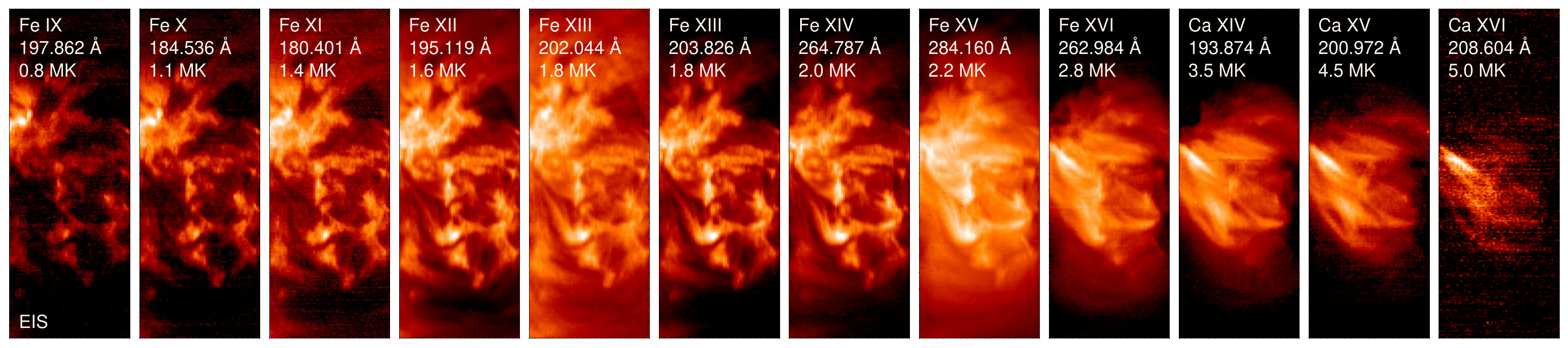}}
  \centerline{
    \includegraphics[angle=90,width=\linewidth]{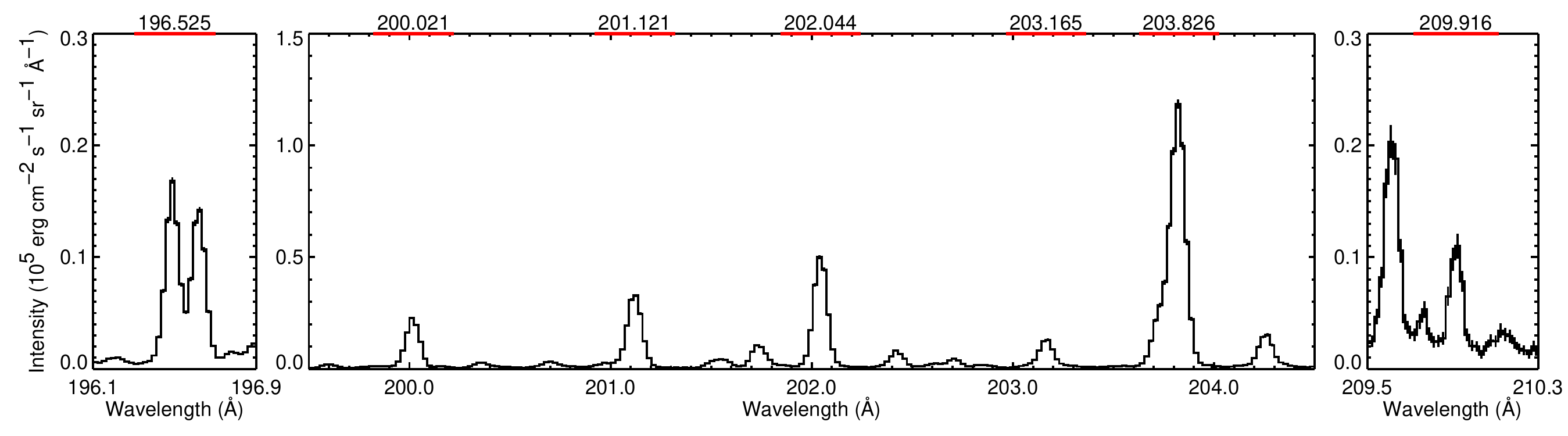}}
  \caption{Observations of NOAA active region 11785 on 8 July 2013 near 2\,UT. The top panels show
    the photospheric magnetic field measured with HMI, million degree emission observed with the
    AIA \ion{Fe}{9} 171\,\AA\ channel, and high temperature loops observed with XRT. The middle
    panels show EIS rasters in a series of emission lines that range in temperature of formation
    from 0.8\,MK to 5.0\,MK. The bottom panel shows an EIS spectrum near 202\,\AA\ from a single
    spatial pixel with the \ion{Fe}{13} lines of interest highlighted. The EIS rasters are from an
    observation that began at 01:55\,UT, and this field of view is indicated by the dotted lines in
    the top panels. The EIS full CCD spectrum is from an observation that began at 00:20\,UT; this
    field of view is indicated by the dashed lines in the top panels.}
  \label{fig:hinode}
\end{figure*}
%% Figure~\ref{fig:hinode}

The problem of interpreting astrophysical spectra is a complex one and it is tempting to account
for any systematic uncertainties in the analysis with broad, ad hoc assumptions. One might assume,
for example, an inflated uncertainty for each observed line intensity with the hope of encompassing
the uncertainty associated with counting statistics, the radiometric calibration, and the atomic
data. Such broad assumptions, however, do not account for obvious correlations within the
analysis. For example, the uncertainty in the calibration clearly should be much smaller for two
emission lines that are close in wavelength than for two lines that are far apart in
wavelength. Similarly, the atomic data for strong lines that are the result of transitions from the
ground state are likely to be more accurate than the atomic data for weak lines that are influenced
by many different transitions \citep[e.g.,][]{foster2010}.

Fortunately, recent reductions in the cost of computing have made it possible to carry out detailed
statistical analysis on complex systems. \citet{lee_cal2011}, for example, have developed a
Bayesian approach for sampling from a distribution of plausible calibration curves and
incorporating this information into a comprehensive analysis of high-energy Chandra spectra.  Along
the same theme, here we consider the effect of uncertainties in the atomic data and the problem of
propagating them to the determination of plasma parameters. Since the atomic calculations involve
many thousands of parameters, we cannot sample the posterior in the usual way.  Thus, we develop a
method that relies on a relatively small number of realizations of the atomic data to compute
posterior distributions for the parameters of interest. While there still exist several
computational bottlenecks in the process, we present for the first time a framework that can be
applied in general to this class of problems. To begin with, we deploy a simple, but realistic,
model to describe the uncertainties in some of the atomic parameters for \ion{Fe}{13}, use it to
generate different realizations of the plasma emissivities, then apply this ensemble of atomic data
to analyze the density-sensitive \ion{Fe}{13} lines observed in solar active region spectra using a
Bayesian framework.

We are focusing here on \ion{Fe}{13} for various reasons: it is one of the most widely used ions;
several atomic calculations are available; the main lines, from the 3s$^2$ 3p 3d configuration, are
well identified and strong in active regions; are mostly free of blends; and fall in a spectral
region where they are observed by the EUV Imaging Spectrometer on \textit{Hinode} (EIS,
\citealt{culhane_eis:07}) and the radiometric calibration is relatively well understood. EIS is
routinely used to measure electron densities from coronal iron ions (e.g., see
\citealt{watanabe_etal:09,young_etal:09_fe_13}). Furthermore, considering lines from a single ion
removes the uncertainties related to the ionization fractions and elemental abundance, greatly
simplifying the problem. Future work will expand this analysis to the calculation of the
undertainties associated with emission measure distributions, which generally use observations of
emission lines from different ionization stages and elements.

This paper is structured in the following way. In Section~\ref{sec:observations} we describe the
\ion{Fe}{13} emission lines of interest and present the traditional analysis of some representative
observations. In Section~\ref{sec:atodat} we develop the models for the uncertainties in the atomic
data and describe how they are used to generate different realizations of the plasma emissivities.
In Section~\ref{sec:test} we present the analysis of a simple test case where we specify the plasma
properties. In Section~\ref{sec:inference} we describe the application to actual observations. In
Section~\ref{sec:discussion} we conclude with a summary and a discussion of future work.

\section{Modeling EIS \ion{Fe}{13} Observations} \label{sec:observations}

To motivate our thinking about the data analysis, we consider the problem of determining the
electron density in loop footpoints using observations of density-sensitive \ion{Fe}{13} emission
lines. The uncertainties in the atomic data and the analysis presented in the next sections are
independent of the specific problem to which they are applied. This analysis could, for example,
also be applied to measuring densities in million degree loops or measuring densities in the
diffuse corona above the limb. We feel, however, that considering a specific application makes the
analysis both realistic and tractable.

Figure~\ref{fig:hinode} illustrates a typical observation of a solar active region. The intense
magnetic fields in the active region lead to the formation of 3--4\,MK plasma on the relatively
short loops in the active region core \citep[e.g.,][]{delzanna2014b,delzanna2014a,warren2012}. The
footpoints of these high temperature loops are bright in million degree emission lines, and the
footpoints are often referred to as the ``moss'' because of their mottled appearance in high
resolution images \citep[e.g.,][]{berger1999,fletcher1999}. These footpoint measurements provide
information on the boundary conditions in these loops and are important for constraining models of
coronal heating \citep[e.g.,][]{peres1994,martens2000,winebarger2008}. Measurements of the electron
density in the moss are of particular utility because they yield information on both the base
pressure of the loop as well as the filling factor \citep{warren2008}.

\begin{figure}[t!]
  \centerline{
    \includegraphics[width=\linewidth]{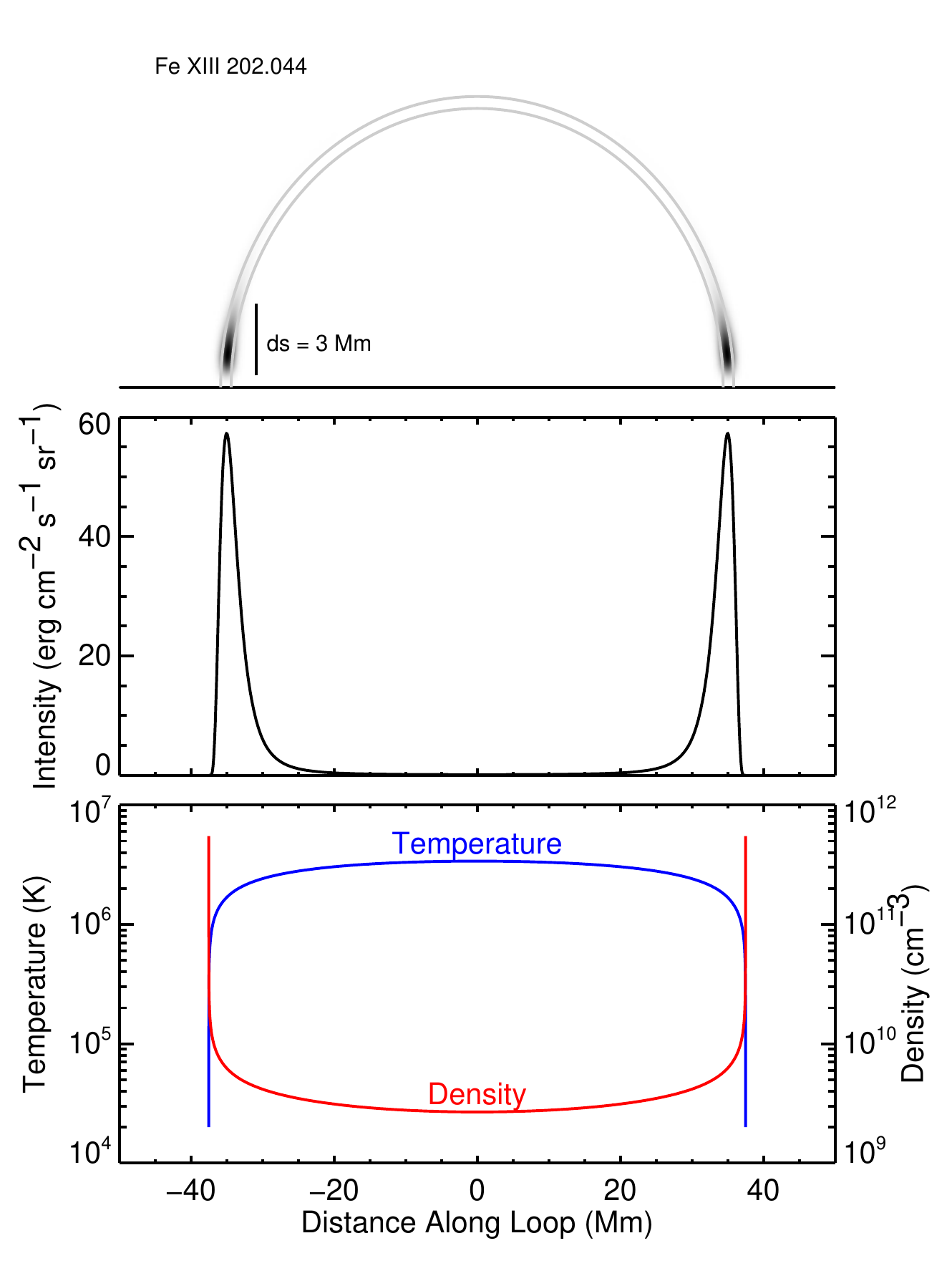}}
  \caption{Density measurements in the moss can be used in conjunction with hydrodynamic loop
    models to infer the properties of high-temperature active region loops. The bottom panel shows
    the temperature and density as a function of position derived from assuming steady, uniform
    heating. The middle panel shows the intensity of \ion{Fe}{13} 202.044\,\AA\ as a function of
    position along the loop. The top panel shows a rendering of the loop as it would appear viewed
    from the side. See \cite{warren2008} for additional details.}
  \label{fig:model}
\end{figure}

To further illustrate the concept of the moss, in Figure~\ref{fig:model} we show \ion{Fe}{13}
202.044\,\AA\ intensities computed from a simple, one-dimensional hydrodynamic loop model
\citep{schrijver2005}. Here a relatively large volumetric heating is assumed and a loop with an
apex temperature of about 3\,MK is produced. As expected, the \ion{Fe}{13} emission comes from a
relatively localized region near the footpoint of the loop. Thus moss observations can yield
information on individual loops. The emission at higher temperatures, in contrast, is generally an
integration across many different loops along the line of sight, making the interpretation of such
observations much more difficult.

\begin{deluxetable}{@{}cr@{ -- }lrcl@{}}
  \tabletypesize{\scriptsize}
  \tablecaption{Selected \ion{Fe}{13} lines observed by \textit{Hinode} EIS \label{tab:lines} }
  \tablehead{
    \\ [-1.5ex]
    \multicolumn{1}{c}{$i$--$j$\tablenotemark{a}} &
    \multicolumn{2}{c}{Identification} &
    \multicolumn{1}{c}{$\lambda_{\rm exp}$ (\AA)} &
    \multicolumn{1}{c}{Notes}
  }
  \startdata
  1--20 & 3s$^2$ 3p$^2$ $^3$P$_{0}$ & 3s$^2$ 3p 3d $^3$P$_{1}$ & 202.044 & \\
  2--23 & 3s$^2$ 3p$^2$ $^3$P$_{1}$ & 3s$^2$ 3p 3d $^3$D$_{1}$ & 201.126 & \\
  3--20 & 3s$^2$ 3p$^2$ $^3$P$_{2}$ & 3s$^2$ 3p 3d $^3$P$_{1}$ & (209.916) & branching ratio \\
  3--25 & 3s$^2$ 3p$^2$ $^3$P$_{2}$ & 3s$^2$ 3p 3d $^3$D$_{3}$ &  203.826 & self-blend \tablenotemark{b} \\
  3--24 & 3s$^2$ 3p$^2$ $^3$P$_{2}$ & 3s$^2$ 3p 3d $^3$D$_{2}$ & 203.795  & self-blend
  \tablenotemark{b} \\
  7--60 &     3s 3p$^3$ $^3$D$_{1}$ & 3s 3p$^2$ 3d $^3$F$_{2}$ &  203.772 & self-blend \tablenotemark{b}\\
  8--60 &     3s 3p$^3$ $^3$D$_{2}$ & 3s 3p$^2$ 3d $^3$F$_{2}$ &  203.835 & self-blend \tablenotemark{b}\\
  3--23 & 3s$^2$ 3p$^2$ $^3$P$_{2}$ & 3s$^2$ 3p 3d $^3$D$_{1}$ &  (204.942) & branching ratio \\
  1--23 & 3s$^2$ 3p$^2$ $^3$P$_{0}$ & 3s$^2$ 3p 3d $^3$D$_{1}$ &  (197.431) & branching ratio \\
  2--24 & 3s$^2$ 3p$^2$ $^3$P$_{1}$ & 3s$^2$ 3p 3d $^3$D$_{2}$ &  200.021  &  \\
  2--19 & 3s$^2$ 3p$^2$ $^3$P$_{1}$ & 3s$^2$ 3p 3d $^3$P$_{2}$ &  209.619  &  \\
  2--22 & 3s$^2$ 3p$^2$ $^3$P$_{1}$ & 3s$^2$ 3p 3d $^3$P$_{0}$ & 203.165 & blended  \\
  4--26 & 3s$^2$ 3p$^2$ $^1$D$_{2}$ & 3s$^2$ 3p 3d $^1$F$_{3}$ &  196.525   \\
  2--21 & 3s$^2$ 3p$^2$ $^3$P$_{1}$ & 3s$^2$ 3p 3d $^1$D$_{2}$ &  (204.262) & blended \\
  \enddata
  \tablenotetext{a}{$i$ and $j$ are the indices of the lower and upper levels in the
    CHIANTI database}
  \tablenotetext{b}{self-blend: multiple lines from the same ion that are close in
    wavelength}
\end{deluxetable}

As mentioned previously, the EIS instrument on \textit{Hinode} observes many emission lines whose
intensities can be combined to form density-sensitive ratios. Figure~\ref{fig:hinode} also shows
the spectral region observed with EIS near 202\,\AA, which is dominated by intense \ion{Fe}{13}
lines. A typical analysis of these lines involves fitting each of the spectral features with
Gaussians to derive the line intensities and their corresponding statistical errors. For this work
we consider the lines at 196.525, 200.021, 201.121, 202.044, 203.165, 203.826, and
209.916\,\AA\ (see Table~\ref{tab:lines}) and fit them with single or multiple Gaussians, as
appropriate. These lines will be discussed extensively in Section~\ref{sec:atodat}. We have
randomly selected 1000 pixels from the EIS observations of the moss shown in
Figure~\ref{fig:hinode} for analysis. Note that the lines at 202.044 and 209.916\,\AA\ originate in
the same upper level and they form a branching ratio that is independent of solar conditions. The
other five lines form density-sensitive ratios with the 202.044\,\AA\ line.

Since it is not obvious how to reconcile all of the individual line ratios, we use a simple
empirical model to interpret the observations. We make the standard assumption that the intensity
of an emission line can be computed with
\begin{equation}
  I_n = \epsilon_n(n_e, T_e)n_e^2\,ds,
  \label{eq:model}
\end{equation}
where $\epsilon_n(n_e, T_e)$ is the plasma emissivity, $n_e$ and $T_e$ are the electron density and
temperature, and $ds$ is the effective path length through the solar atmosphere (see, for example,
\citealt{mariska1992}). First, we note that since the excitation energies of the levels associated
with these wavelengths are very similar, the emissivity ratios used to evaluate the plasma
densities are highly insensitive to changes in temperature.  Furthermore, as illustrated in
Figure~\ref{fig:model}, most of the \ion{Fe}{13} emission in high-temperature loops are thought to
originate in a narrow region near the footpoint over which the temperatures are close to the peak
temperature of formation of the ion, $\approx{1.8}$~MK.  We therefore adopt this value of the
temperature and henceforth treat the plasma as isothermal.  Of course, hydrodynamic models show
that there are gradients in temperature and density along the loop, where segments of high $n_e$
occupy small $ds$, and segments of small $n_e$ cover a large $ds$, so Equation~\ref{eq:model} must
be treated as an empirical description characterized by a representative density and effective path
length.

With this empirical description, however, we can derive information about the solar atmosphere
directly from the observations. The physical model shown in Figure~\ref{fig:model} depends on
additional assumptions about the loop geometry, the plasma composition, and the nature of the
heating.

%% RAW DATA FROM fe_13_fit_intensities.pro
%%      n chianti = 0
%%         n pixel = 216
%%     model log_n = 9.68 +- 0.010
%%    model log_ds = 8.67 +- 0.021
%%            chi2 = 154.5
%% normalized chi2 = 30.9
%%      Line      Iobs    SigmaI    Imodel      dI/I  dI/Sigma
%%   196.525    1473.1      18.8    1473.8       0.0       0.0
%%   200.021    1521.4      29.1    1749.9      15.0       7.9
%%   201.121    2373.2      44.4    1987.0      16.3       8.7
%%   202.044    2866.5      53.6    2989.1       4.3       2.3
%%   203.165     775.2      42.5     767.5       1.0       0.2
%%   203.826    9237.6     142.6    8751.2       5.3       3.4
%%   209.916     530.2      56.4     516.2       2.6       0.2

\begin{deluxetable}{rr@{ $\pm$ }rrr}
  \tabletypesize{\scriptsize}
  \tablewidth{3in}
  \tablecaption{Modeling \ion{Fe}{13} Line Intensities in the
    Moss\tablenotemark{a}   \label{tab:model}}
  \tablehead{
    \\ [-2.5ex]
    \colhead{Line} &
    \colhead{$I_{obs}$} &
    \colhead{$\sigma_{I}$} &
    \colhead{$I_{model}$} &
    \colhead{$|\Delta I|/I$(\%)}
  }
  \startdata
  196.525 &   1473.1  &    18.8 &   1443.6  &     2.0 \\
  200.021 &   1521.4  &    29.1 &   1749.9  &    15.0 \\
  201.121 &   2373.2  &    44.4 &   1987.0  &    16.3 \\
  202.044 &   2866.5  &    53.6 &   2989.1  &     4.3 \\
  203.165 &    775.2  &    42.5 &    767.5  &     1.0 \\
  203.826 &   9237.6  &   142.6 &   8751.2  &     5.3 \\
  209.916 &    530.2  &    56.4 &    516.2  &     2.6 \\
  \enddata
  \tablenotetext{a}{Observed and modeled line intensities for a single spatial pixel (\#217) using
    Equation~\ref{eq:model}. The best-fit density and path length are $\log n_e = 9.68\pm0.01$ and
    $\log ds = 8.67\pm0.02$. The intensities and their corresponding uncertainties are in units of
    erg cm$^{-2}$ s$^{-1}$ sr$^{-1}$.}
\end{deluxetable}

For this empirical model we can use the observed intensities and their corresponding statistical
uncertainties, the computed plasma emissivities, and assumed temperature to infer the best-fit
electron density and path length by performing a least-squares fit. The plasma emissivites for each
line is computed using version 8 of the CHIANTI atomic data base
\citep{delzanna_etal:2015_chianti_v8,dere1997}.

The results of an example calculation are shown in Table~\ref{tab:model}, where we have taken the
observed intensities from a single spectral pixel (arbitrarily chosen as \#217) from an EIS
full-CCD observation (EIS file {\footnotesize\verb+eis_l0_20130708_002042+}) and applied
Equation~\ref{eq:model}. The resulting best-fit parameters are $\log n_e = 9.68\pm0.01$ (cm$^{-3}$)
and $\log ds = 8.67\pm0.02$ (cm). The error bars associated with the parameters are very small,
suggesting that the parameters are very precisely determined. The uncertainties associated with the
intensities, however, are also small and the standard method of determining the best-fit $\Ne$ and
$\ds$ by minimizing $\chi^2$ results in reduced $\chi^2$$\approx$$30$ for this case, indicating
that the model is a poor fit to the data.

This example highlights the difficulty in interpreting many solar observations. Since the sun is
relatively close, we can obtain observations with high signal-to-noise. This bounty of photons,
however, means that models generally do not pass rigorous statistical tests. This can be simply
ignored or covered up by inflating the statistical errors with ad hoc assumptions. The real
deficiency in the analysis is taking the atomic data as fixed and without uncertainty. In reality,
the uncertainties associated with the plasma emissivities are likely to be comparable to or larger
than those from counting statistics, and a proper data analysis must include a treatment of
them. We now turn to estimating the uncertainties in the atomic data available for \ion{Fe}{13}.

\section{Uncertainties in the Atomic Data} \label{sec:atodat}

The most recent (and largest) scattering calculation for \ion{Fe}{13} is an $R$-matrix calculation
carried out within the UK APAP network\footnote{www.apap-network.org}, which had a target of 749
levels up to $n=4$ \citep{delzanna_storey:2012_fe_13}.  The main focus of this calculation was to
provide accurate data for the $n=4 \to n=3$ soft X-ray transitions. Indeed, new lines in this
wavelength range were subsequently identified \citep{delzanna:12_sxr1}.  The scattering calculation
was supplemented by a structure calculation which was used to calculate the radiative data, using
either observed or empirically-adjusted theoretical wavelengths.  The scattering and radiative data
produced in this calculation were recently made available within the CHIANTI
database\footnote{www.chiantidatabase.org} in its version 8 \citep{delzanna_etal:2015_chianti_v8}.
We use these data as our baseline.

\citet{storey_zeippen:2010} previously carried out a similar scattering calculation (using the same
$R$-matrix method and the same codes), the only difference being that it was aimed at improving the
earlier calculations for the $n=3$ levels.  The target had a total of 114 fine-structure levels,
and included only some $n=4$ levels.

\citet{delzanna_storey:2012_fe_13} also performed separate calculations for the $n=5,6$ levels, but
showed that cascading effects are small when considering the strong EUV lines emitted by the $n=3$
levels.  The same paper also showed that the intensities of the transitions from the $n=3$ levels
are close to those of the previous \cite{storey_zeippen:2010} model.

The \cite{storey_zeippen:2010} atomic data provided very good agreement between observed and
theoretical intensities of the strongest EUV lines, as shown in one of the benchmark works by
\cite{delzanna:11_fe_13}, based on a variety of sources, including Hinode/EIS.
\cite{delzanna:11_fe_13} also benchmarked other atomic data for this ion, calculated by
\cite{gupta_tayal:98} and \cite{aggarwal_keenan:05}.  Various shortcomings in these calculations
were found.  On the other hand, excellent agreement (to within a relative 10\%) was found for the
main lines observed by Hinode EIS in an active region moss area and the \cite{storey_zeippen:2010}
atomic data, already indicating an excellent accuracy in both the experimental and theoretical
data, as we will also confirm below.

The \cite{delzanna:11_fe_13} benchmark work also reviewed all the previous identifications and
determined which lines are likely to be blended, and hence avoided in our analysis. The main lines
chosen for our study are listed in Table~\ref{tab:lines}, in order of decreasing intensity.  The
main line is the straight decay to the ground state, which produces the emission line at
202.044~\AA.  We note that two transitions from the 3s 3p$^2$ 3d $^3$F$_{2}$ were suggested to be
blending the main density diagnostic line, already a self-blend at 203.8~\AA. However, as noted by
\cite{delzanna:11_fe_13}, at the high densities that are considered here these lines do not have a
significant contribution, so even if the identifications were incorrect the results presented here
would still stand.  The other lines (at 201.126, 200.021, 209.619, 203.165~\AA) have a similar
density sensitivity as the 203.8~\AA\ one, except the 196.525~\AA, which decays to a more excited
level.  The 203.165~\AA\ was shown to be blended at low densities.  Some lines (with wavelengths in
brackets) were not considered since they are branching ratios (decays from the same upper level)
with other lines we have included. Further benchmarks were carried out by \cite{delzanna:12_atlas}.

\begin{figure}[t!]
  \centerline{
    \includegraphics[clip, width=0.9\linewidth]{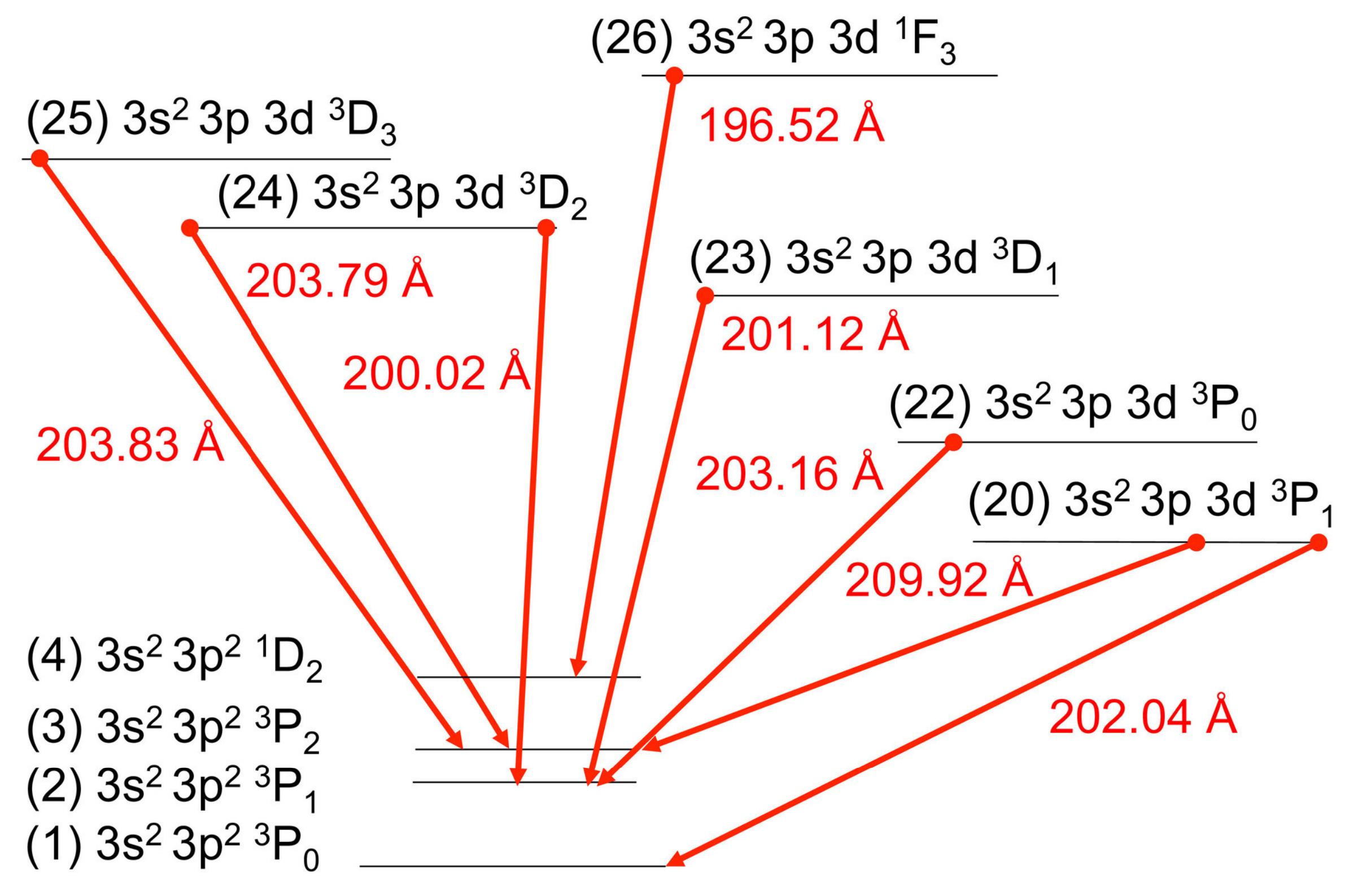}}
  \caption{A simplified level diagram for the transitions relevant to the lines considered in this paper.}
  \label{fig:levels}
\end{figure}

\begin{figure}[!htbp]
  \centerline{
    \includegraphics[width=0.95\linewidth]{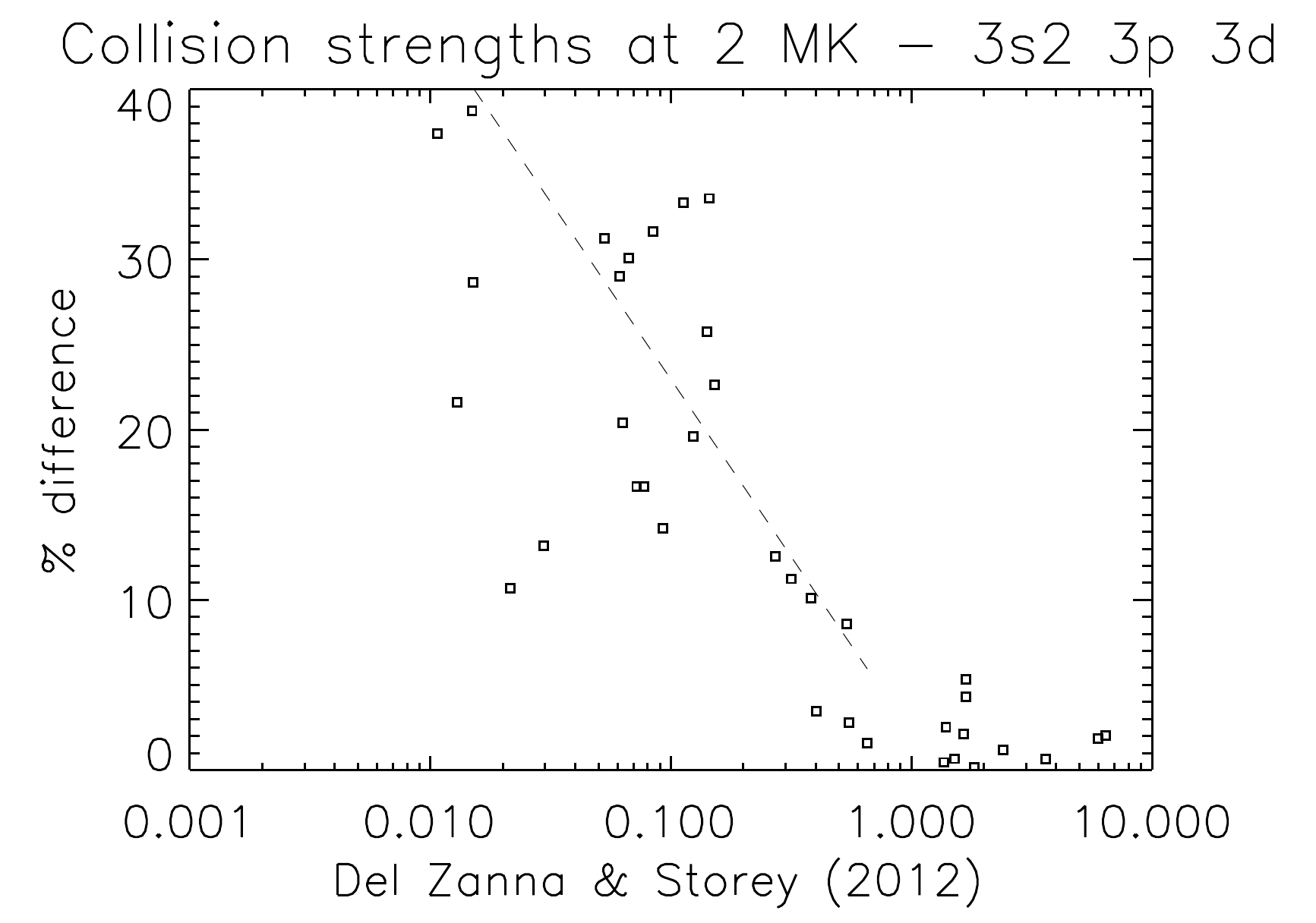}}
  \centerline{
    \includegraphics[width=0.95\linewidth]{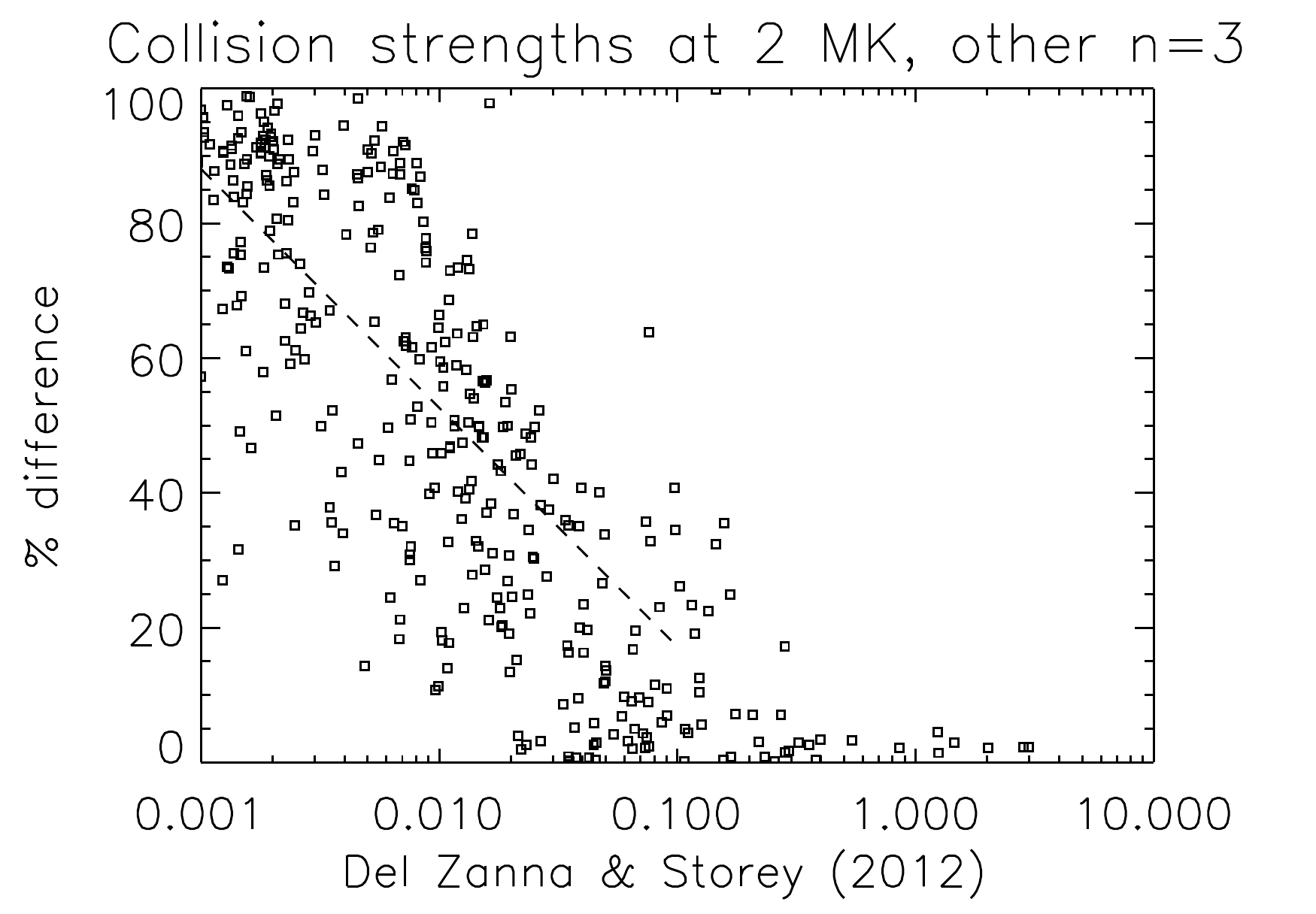}}
  \caption{Percentage difference in the thermally-averaged collision strengths
    (\cite{storey_zeippen:2010} vs.\ \cite{delzanna_storey:2012_fe_13}), for a selection of
    transitions (top: to the 3s$^2$ 3p 3d levels; bottom: to the other $n=3$ levels). The dashed
    lines represent the approximate uncertainties used to generate alternative realizations of
    the atomic data for \ion{Fe}{13}.}
  \label{fig:comp_ups}
\end{figure}
%% Figure~\ref{fig:comp_ups}

The intensity of a spectral line is proportional to the population of the upper level and the
spontaneous transition probability (the A-value). To assess which atomic rates affect a spectral
line, it is therefore important to check which are the main populating mechanisms for each level. A
simplified level diagram for the main transitions discussed in this section is provided in Figure
3, for a specific density.  However, the main populating mechanisms for each atomic level normally
vary with the density, so the issue can become quite complex to describe.  Some details will be
discussed in a separate paper, where also the various parameters that can affect an atomic
calculation are reviewed.  In summary, the 202.044~\AA\ line is mainly populated by direct
excitation from the 3s$^2$ 3p$^2$ $^3$P$_{0}$ ground state via a strong dipole-allowed
transition. The various calculations provide the same rate, within a few percent.  In turn, the
population of the ground state decreases significantly as the population of the metastable levels
increases.  On the other hand, the populations of the other levels which produce the other lines in
Table~\ref{tab:lines} are mainly driven by excitations from all the 3s$^2$ 3p$^2$ $^3$P$_{0,1,2}$
levels, although non-negligible contributions (typically 10--30\%) come also from cascading from
higher levels.

\begin{figure}[!htbp]
  \centerline{
    \includegraphics[width=0.95\linewidth]{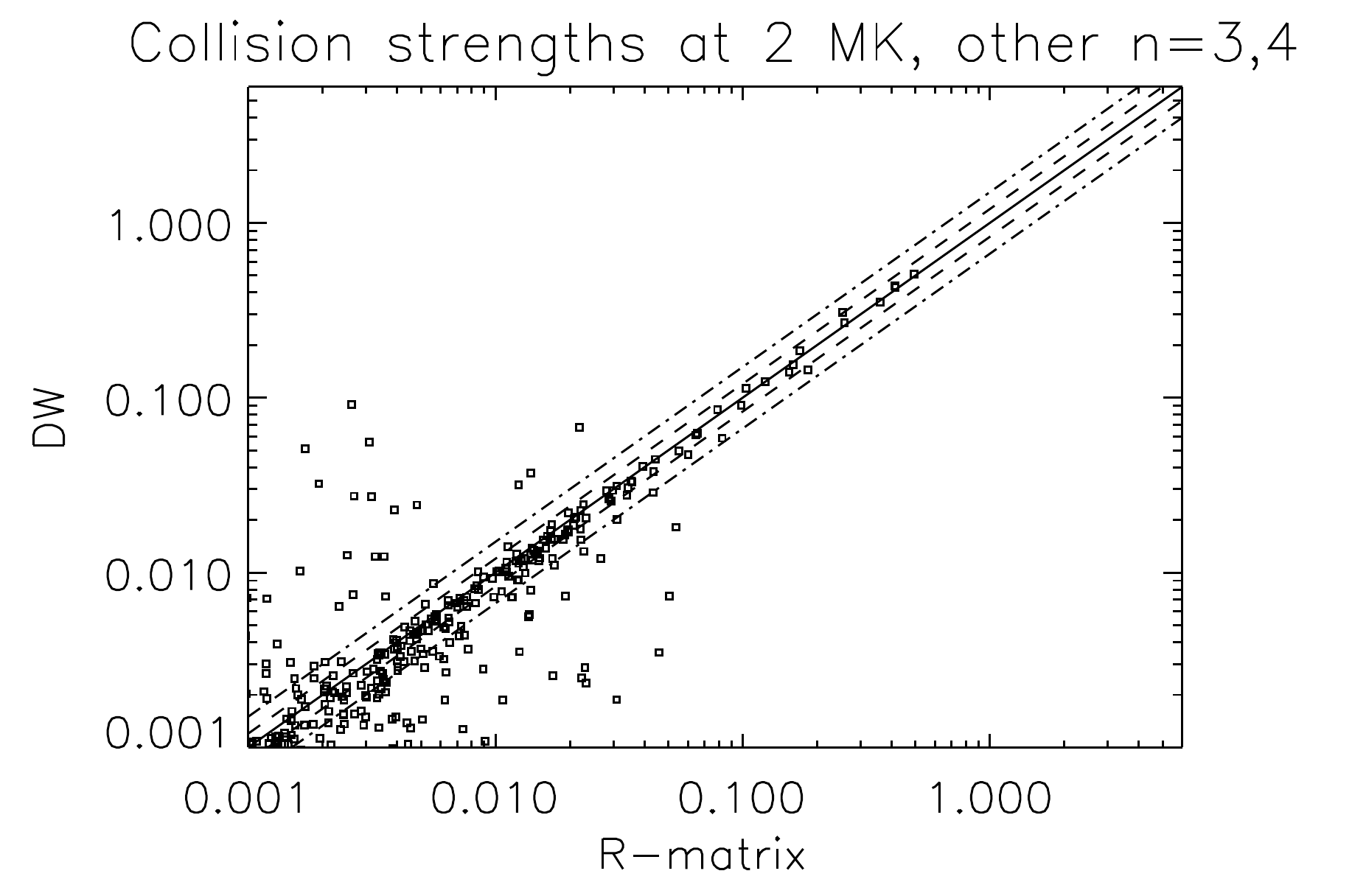}}
  \caption{Scatter plot of the thermally-averaged collision strengths calculated by
    \cite{delzanna_storey:2012_fe_13} with the $R$-matrix codes (as in CHIANTI, present), vs.\ with
    the DW approximation. Dash and dot-dash lines indicate a $\pm$ 20, 50\%.}
  \label{fig:comp_ups2}
\end{figure}
%% Figure~\ref{fig:comp_ups2}

It is therefore important to first assess how accurate the rates of excitation from the 3s$^2$
3p$^2$ $^3$P$_{0,1,2}$ to the 3s$^2$ 3p 3d levels are.  We have chosen to compare the latest values
with those calculated by \cite{storey_zeippen:2010}, because the two calculations were very
similar, i.e.\ the main differences are caused by the size of the target and not by the method of
the calculation. As already shown by \cite{delzanna_storey:2012_fe_13}, the largest calculation
provides very similar rates for the stronger lines, but significantly increased values for the
weaker ones, as one would expect.  We have considered only excitations from the 3s$^2$ 3p$^2$
$^3$P$_{J}$ and 3s$^2$ 3p$^2$ $^1$D$_{2}$ levels (the only ones with significant population) at the
temperature of peak ion abundance in ionization equilibrium (2 MK, see Figure~\ref{fig:comp_ups}
top).  As an estimate of the uncertainty in the strongest lines, with collision strengths above
1.0, we have taken 5\%, which is well above the scatter of values.  For the weaker lines, we have
taken as an estimate the dashed line, i.e.  a linear increase (up to a maximum of 50\%).

We have then considered all the excitations to the remainder of the $n=3$ levels calculated by
\cite{storey_zeippen:2010}, taking into account the different level orderings of the two
calculations.  In this case, we have taken a 10\% uncertainty for the transitions above 0.1, and
the linear increase shown in Figure~\ref{fig:comp_ups} (bottom, up to a maximum of 50\%).

One possible estimate for all the $n=3$ levels not included in \cite{storey_zeippen:2010} and all
the $n=4$ levels is to compare the full scattering calculation with the results of the
distorted-wave (DW) calculation carried out by \cite{delzanna_storey:2012_fe_13}, which does not
include resonance enhancements (see Figure~\ref{fig:comp_ups2}).  We have taken a 20\% uncertainty
for the transitions above 0.01 and 50\% for the weaker transitions.

\begin{figure}[!htbp]
  \centerline{
    \includegraphics[width=0.95\linewidth]{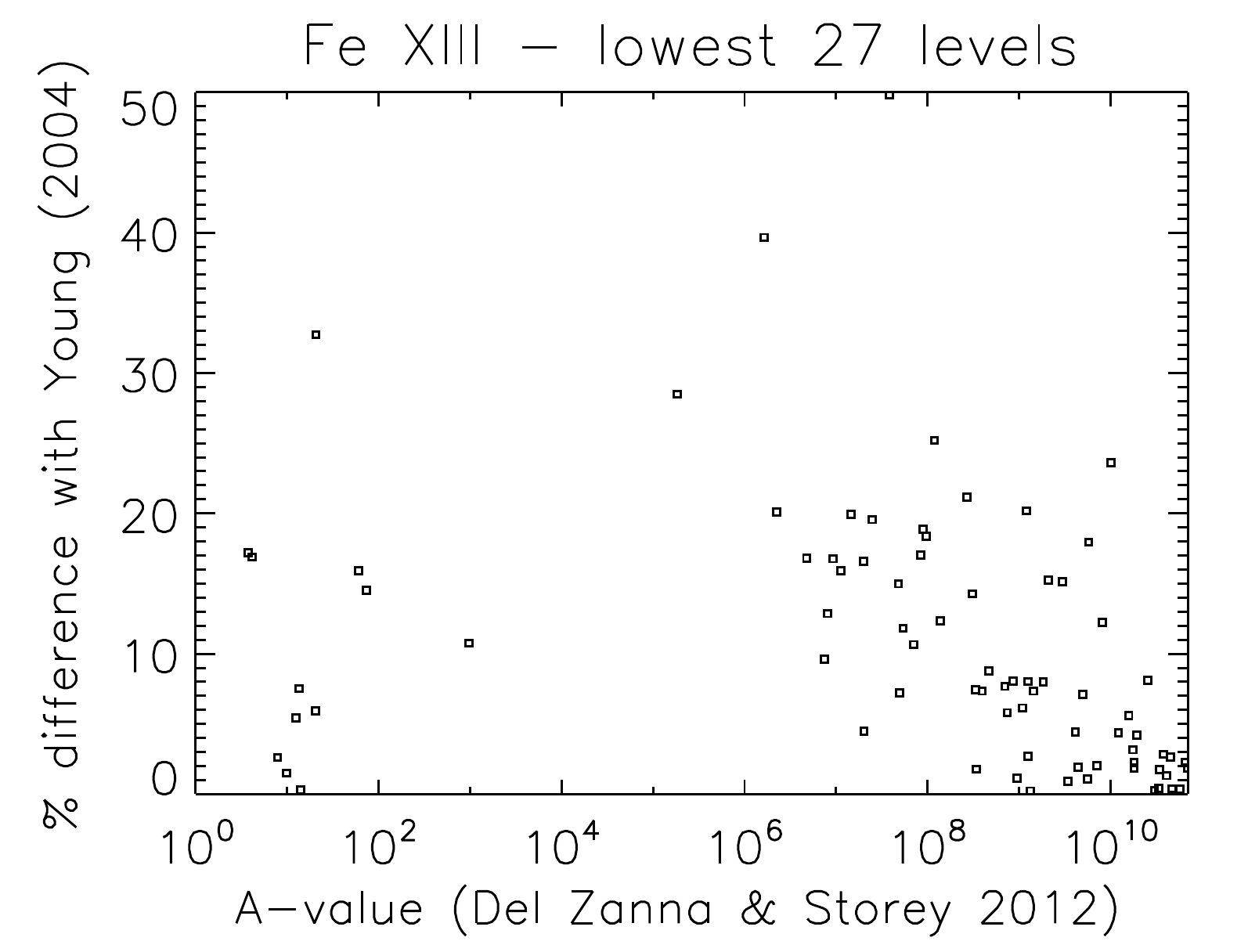}}
  \caption{Percentage difference in the A-values calculated by \cite{young:04_fe_13} and
    \cite{delzanna_storey:2012_fe_13} within the lowest 27 levels.}
  \label{fig:comp_a}
\end{figure}
% Figure~\ref{fig:comp_a}

The next step is to provide an estimate on the uncertainty of the A-values. As shown by
\cite{young:04_fe_13}, different calculations can provide significantly different values. For our
estimates, we have chosen to compare the \cite{delzanna_storey:2012_fe_13} A-values with those
calculated by \cite{young:04_fe_13} with the SUPERSTRUCTURE program \citep{eissner_etal:74}.  An
extended configuration set was used by \cite{young:04_fe_13} to calculate radiative data for this
ion.  This data were made available within CHIANTI version 4 \citep{young_etal:03_chianti_v4} in
2003.  Figure~\ref{fig:comp_a} shows comparisons of the A-values for all the transitions within the
lowest 27 levels, which include the 3s$^2$ 3p 3d.

We have taken for transitions having an A-value above 10$^{10}$ an uncertainty of 5\%, for those
between 10$^{8}$ and 10$^{10}$ 10\%, while for weaker transitions, 30\%.  For the forbidden
transitions within the ground configuration: we have taken 10\%.

\begin{figure*}[!htbp]
  \centerline{
    \includegraphics[angle=90,width=0.975\linewidth]{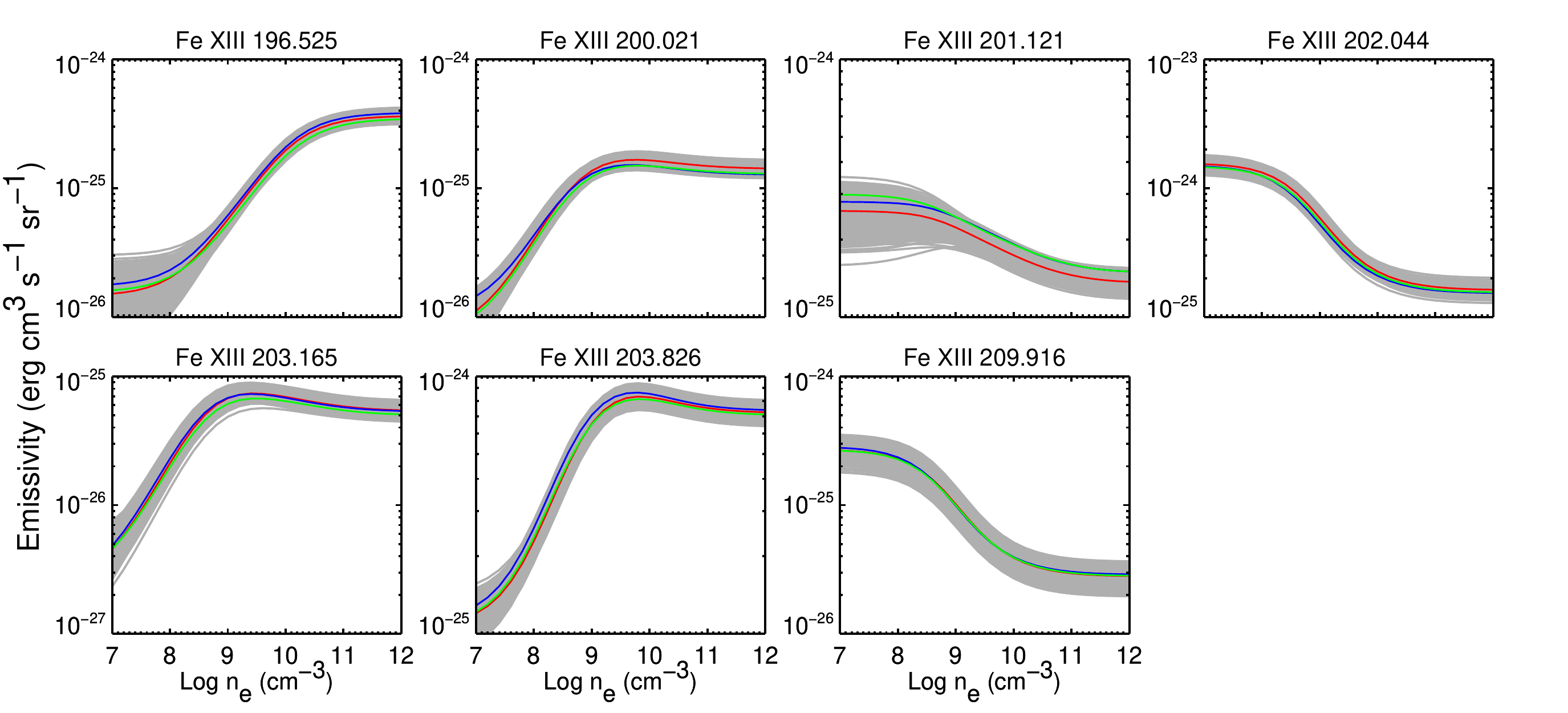}}
  \caption{Emissivities of the seven \ion{Fe}{13} lines considered in this work. The emissivities
    are computed assuming a temperature of 1.8\,MK, the temperature of formation for
    \ion{Fe}{13}. The gray lines represent the 1000 realizations of the CHIANTI atomic data. The
    red curve is the default value from CHIANTI v.8. As discussed in Section~5, the
      blue curve is identified as being most probable match to the observations (\#471) and the
      green curve the second most probable (\#368).}
  \label{fig:emiss}
\end{figure*}
%% Figure~\ref{fig:emiss}

\begin{figure*}[!htbp]
  \centerline{
    \includegraphics[angle=90,width=0.975\linewidth]{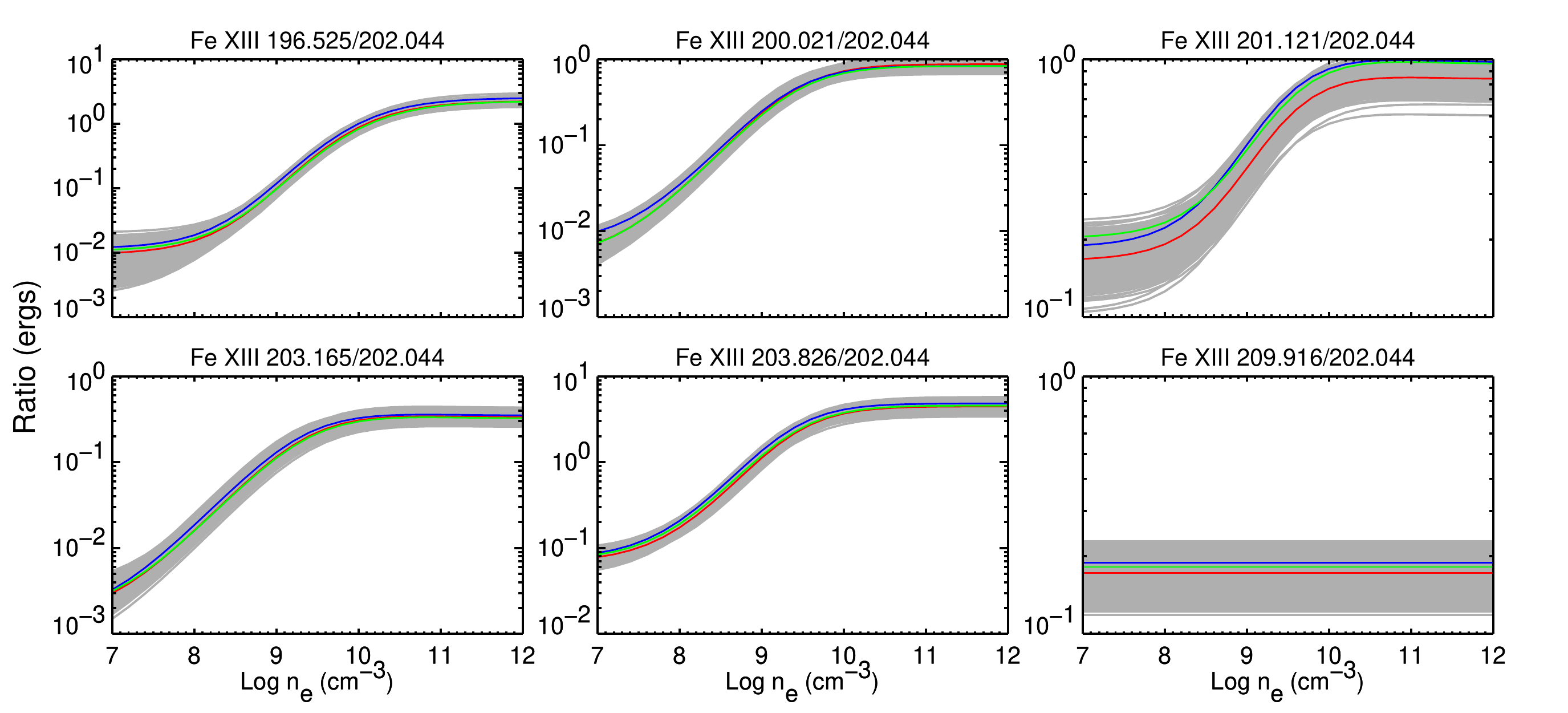}}
  \caption{The theoretical ratios of six \ion{Fe}{13} lines with respect to \ion{Fe}{13} 202.044.
  See Figure~\ref{fig:emiss} for additional details.}
  \label{fig:ratios}
\end{figure*}
%% Figure~\ref{fig:ratios}

We have modified the standard CHIANTI IDL routines distributed in SolarSoft (SSW,
\citealt{freeland1998}) to assign to each transition an uncertainty in the A-value and in the
excitation rate.  We used the IDL function {\tt randomn} to randomly vary each rate within the
estimated uncertainty. The distribution is normal, in the sense that if, e.g., an uncertainty is
10\%, most values will vary within $\pm$ 20\%.  We then used the standard CHIANTI routine ({\tt
  emiss\_calc}) to calculate the line emissivities.  For each of the seven chosen lines, we have
added any \ion{Fe}{13} lines within $\pm$ 0.1~\AA\ to take self-blends into account.  We have
generated a total of 1000 realizations of the emissivities for each line, which are shown in
Figure~\ref{fig:emiss}.  The figure clearly shows how the spectral lines vary their emissivities as
a function of the density.  Figure~\ref{fig:ratios} shows the variation of ratios with density.

By attaching reasonable uncertainties to the atomic data we can generate realizations of the
emissivities that capture this uncertainty. We can then use the ensemble of emissivities to
characterize parameters like plasma densities and column heights.  Our methodology for combining an
ensemble of emissivities with observed data to account for uncertainties in atomic data is
described in detail in Section~\ref{sec:inference}.
%\section{Simulated Pixels} \label{sec:test}
\section{Simulated Datasets} \label{sec:test}

The next step in this analysis is to generate sets of intensities from known densities and path
lengths. This will allow us to test our ability to recover physical parameters from the
\ion{Fe}{13} intensities and to illustrate how the variations in the atomic data developed in the
previous section lead to variations in the inferred densities and path lengths. These intensities
complement the set of observed intensities taken from the data illustrated in
Figure~\ref{fig:hinode}.

The ratio curves shown in Figure~\ref{fig:ratios} indicate that these lines are sensitive to
density in the range of $n_e$ of $10^8$ to $10^{11}$\,cm$^{-3}$.  Thus we randomly select 1000
densities uniformly on the interval $\log n_e = [8.5, 10.5]$. To continue with our theme of
analyzing observations of active region moss we use the theoretical estimate of the moss path
length from \citet{martens2000} of
\begin{equation}
  ds = \frac{2.56\times10^8}{P_0}, \label{eq:ds}
\end{equation}
where $P_0=2k_bn_eT_e$ is in dyne cm$^{-2}$, $k_b$ is the Boltzmann constant, and $ds$ is in
cm. Again, we use the peak temperature of formation for \ion{Fe}{13}, 1.8\,MK, for this
calculation. Note that this expression was derived for somewhat cooler emission and we do not
expect it to track the \ion{Fe}{13} path lengths exactly.

Each set of density and path length can be used to generate a set of intensities using statistics
using the EIS pre-flight effective areas and the assumption of a 60\,s exposure time and the
2\arcsec\ slit. Finally, we do not use these computed intensities directly when attempting to
recover them with the model. We first apply a normally distributed random perturbation to each
intensity, which mimics the variations in measured counts expected due to the finite exposure time.

%% RAW DTA
%%
%%  model log_n = 9.90 +- 0.012  [9.90]
%% model log_ds = 7.82 +- 0.025  [7.82]
%%         chi2 = 4.0
%% normalized chi2 = 0.8
%%      Line      Iobs    SigmaI    Imodel      dI/I  dI/Sigma
%%   196.525     751.0       9.4     750.7       0.0       0.0
%%   200.021     678.7      11.8     694.7       2.4       1.4
%%   201.121     750.7      14.8     748.6       0.3       0.1
%%   202.044    1012.5      20.6    1011.9       0.1       0.0
%%   203.165     313.7      13.7     296.2       5.6       1.3
%%   203.826    3533.1      52.2    3498.8       1.0       0.7
%%   209.916     168.4      21.0     174.8       3.8       0.3
%%
\begin{deluxetable}{rr@{ $\pm$ }rrr}
  \tabletypesize{\scriptsize}
  \tablewidth{3in}
  \tablecaption{Modeling \ion{Fe}{13} Line Intensities in the Moss\tablenotemark{a} \label{tab:test1}}
  \tablehead{
    \\ [-1.5ex]
    \multicolumn{1}{c}{Line} &
    \multicolumn{1}{c}{$I_{obs}$} &
    \multicolumn{1}{c}{$\sigma_{I}$} &
    \multicolumn{1}{c}{$I_{model}$} &
    \multicolumn{1}{c}{$|\Delta I|/I$(\%)}
  }
  \startdata
  196.525  &   751.0  &     9.4 &    750.7  &     0.0  \\
  200.021  &   678.7  &    11.8 &    694.7  &     2.4  \\
  201.121  &   750.7  &    14.8 &    748.6  &     0.3  \\
  202.044  &  1012.5  &    20.6 &   1011.9  &     0.1  \\
  203.165  &   313.7  &    13.7 &    296.2  &     5.6  \\
  203.826  &  3533.1  &    52.2 &   3498.8  &     1.0  \\
  209.916  &   168.4  &    21.0 &    174.8  &     3.8
  \enddata
  \tablenotetext{a}{An example set of intensities computed from assumed values of $\log n_e = 9.90$
    and $\log ds = 7.82$. The observed intensities include a normally distributed random
    perturbation. The best-fit density and path length are $\log n_e = 9.90\pm0.01$ and $\log ds =
    7.82\pm0.03$. The format of the table is the same as Table~\ref{tab:model}.}
\end{deluxetable}

An example set of simulated intensities is given in Table~\ref{tab:test1} for assumed values of
$\log n_e = 9.90$ and $\log ds = 7.80$. If we use the standard set of CHIANTI emissivities to model
these intensities we recover the input parameters almost exactly, $\log n_e = 9.90\pm0.01$ and
$\log ds = 7.82\pm0.03$. As with the example set of observed intensities in Table~\ref{tab:model},
the uncertainties on these parameters are very small. Unlike the case with the observed
intensities, however, we obtain a reduced $\chi^2$ of order 1.
\section{Inference} \label{sec:inference}

The standard method of $\chi^2$ minimization (see Section~\ref{sec:observations}) allows best-fit
values of the plasma density $n_e$ and column depth $ds$ to be determined for a given pixel, under
the assumption that the emissivity curves are completely and correctly specified.

Now, equipped with $1000$ datasets corresponding to randomly selected EIS pixels, we can consider
the uncertainties in the fitted density and path length in each case that result from both
statistical fluctuations in the observed intensities and the atomic data uncertainties incorporated
in the ensemble of CHIANTI emissivities. To do so, we develop a Bayesian methodology that first
assumes the observed data is uninformative regarding the atomic physics (the so-called
\textit{pragmatic Bayesian} method) and then incorporate the potential information in the observed
data to learn about the atomic physics (the \textit{fully Bayesian} method). The fully Bayesian
method is a principled statistical analysis, while the pragmatic method makes simplifying
assumptions that tend to overestimate the final uncertainty on the fitted density and path
length. More details of the distinction between the two methods is discussed in
Section~\ref{sec:posterior}. We start by providing an introduction to Bayesian inference in
Section~\ref{sec:Bayes}.

\subsection{Bayesian Inference} \label{sec:Bayes}

We take a Bayesian approach in our statistical analysis because it enables us to build in the
complex hierarchical dependencies engendered by atomic uncertainties. Such an approach offers a
probability-based formalism for combining information from our prior knowledge and the current
data. This requires both a \textit{prior distribution}, which quantifies the uncertainty in the
values of the unknown model parameters before the data is observed, and a \textit{likelihood
  function} --- the distribution of the data given the model parameters. The likelihood function
allows us to assess the viability of a parameter value given the observed data under a proposed
statistical model. The likelihood function is combined with the prior distribution to yield the
\textit{posterior distribution}, which quantifies the uncertainty in the values of the unknown
model parameters taking account of the observed data. If we let $X$ and $\psi$ represent generic
data and unknown model parameters, respectively, Bayes' theorem provides the posterior distribution
as
\begin{equation}
  p(\psi | X) = \frac{L(\psi | X) \ p(\psi)}{p(X)},
  \label{eq:Bayes}
\end{equation}
where $L(\psi | X)$ is the likelihood of $X$ given $\psi$ (sometimes written as $p(X|\psi)$) and
$P(\psi)$ the prior distribution of $\psi$. The term $P(X)$ is a normalizing constant necessary to
make $p(\psi | X)$ a proper probability distribution. (The term $p(X)$ is sometimes referred to as
the "evidence" in the astrophysics literature.) The posterior distribution, which combines
information in the data with our prior knowledge, is our primary statistical tool for deriving
parameter estimates and their uncertainties.

To perform a Bayesian analysis of the Fe XIII intensities, we start by defining notation and
terminology in Section~\ref{sec:Data}. We specify the likelihood function and the prior
distribution in Section~\ref{sec:statisticalmodel}.  In Section~\ref{sec:posterior}, we derive the
posterior distribution under two sets of assumptions, which result in the aforementioned pragmatic
Bayesian and fully Bayesian approaches. In Section~\ref{sec:separateanalysis} and
Section~\ref{sec:jointanalysis} we discuss our model-fitting routines, separate pixel-by-pixel and
simultaneous analyses, where we consider the $1000$ pixel datasets individually and
simultaneously. In Section~\ref{sec:2stepMCwithMH_testset} and
Section~\ref{sec:2stepMCwithMH_observedset} we apply our methodologies to the simulated and the
observed intensities, respectively.

\subsection{Notation} \label{sec:Data}

Suppose that in each of $K=1000$ pixels we observe the intensities of each of $J$ spectral lines
with wavelengths $\Lambda = \{\lambda_1, \ldots, \lambda_J\}$. Let $I_{k \lambda}$ be the observed
intensity of the line with wavelength $\lambda \in \Lambda$ in pixel $k \in \{1, \ldots, K\}$,
$\sigma_{k \lambda}$ its known standard deviation, $\datak = (I_{k \lambda_1}, \ldots, I_{k
  \lambda_J})$, and $\Data = \{\data_1, \ldots, \data_K\}$.

We also have a collection of $M=1000$ realizations of the plasma emissivities, denoted by
$\mathcal{M}$,
\[
\mathcal{M} = \{ \emisk, \lambda \in \Lambda, m = 1,\ldots,M \},
\]
where $\Nek$ and $\Tek$ are the electron density and temperature for pixel $k$ and $\m$ indexes the
emissivity realization (i.e., emissivity curve, $\emisk$), with $m$=$1$ corresponding to the
  default CHIANTI emissivities.

The expected intensity of the line with wavelength $\lambda$ in pixel $k$ can be rewritten (from
Eq~(\ref{eq:model})) as $\emisNOm \Nek^2 \dsk$, where $\dsk$ is the path length through the solar
atmosphere for pixel $k$. Let $\paramk = (\log \Nek, \log \dsk)$ be the plasma parameters in pixel
$k$, and $\paramj = (\param_1, \ldots, \param_K)$.

\subsection{Statistical Model} \label{sec:statisticalmodel}

The first step in specifying our statistical model is to construct the likelihood function. We
model the intensities $I_{k \lambda}$ given $\rvm$, $\Nek$, and $\dsk$ as a normal (i.e. Gaussian)
distribution,
\begin{equation}
  I_{k \lambda} \mid \rvm, \Nek, \dsk \distas{\rm{indep}} \calN\left(\emisk \Nek^2 \dsk, \,
  \sigma^2_{k \lambda}\right),
\end{equation}
for $\lambda \in \Lambda$, where $\calN(\mu, \sigma^2)$ is a normal distribution with mean $\mu$
and variance $\sigma^2$. We suppress the conditioning on the $\sigma_{k \lambda}$ throughout for
notational simplicity.  Thus the likelihood function of $\datak$ given emissivity index, $\rvm$,
and plasma parameters, $\paramk$, is
\begin{equation}
  \begin{split}
    L(\rvm, \paramk \mid \datak) & = p(\datak \mid \rvm, \paramk) \\ & = \prod_{j=1}^J \calN
    \left(\left.I_{k \lambda_j} \ \right| \ \emiskj \Nek^2 \dsk, \ \sigma^2_{k \lambda_j} \right),
  \end{split}
  \label{eq:likelihood}
\end{equation}
where $\calN(x \mid \mu, \sigma^2)$ is the density of a normal distribution with mean $\mu$ and
variance $\sigma^2$ evaluated at $x$. Note that we focus on methods that treat the emissivity index
$\rvm$ as an unknown parameter, whose prior is specified below, whose posterior we estimate to
determine the most likely emissivity realizations among those in $\mathcal{M}$, and whose
uncertainties affect both the fit and error bars of $\paramk$.

Next, we specify the joint prior distribution on the unknown model parameters. For $\log \Nek$ and
$\rvm$ we specify a continuous uniform distribution and a discrete uniform distribution,
respectively,
\begin{align}
  p(\log \Nek) &= \frac{1}{5} \quad \ \ \text{for } 7 \le \log \Nek \le 12, \label{eq:prior_logne}
  \\ p(m) &= \frac{1}{M} \quad \text{for each } m = 1,\ldots,M. \label{eq:prior_m}
\end{align}
This choice of prior on $\rvm$ stipulates that the $1000$ realizations of emissivity curves in
$\mathcal{M}$ are all a priori equally likely to be the true emissivity. As the realizations were
generated by attaching reasonable uncertainties to the atomic data as described in
Section~\ref{sec:atodat}, the atomic data uncertainties are contained in $p(\rvm)$ and are thus
captured by the corresponding posterior distribution. Therefore, the $1000$ realizations of
emissivity curves can also be considered as a sample of $1000$ draws from an implicit prior
distribution.

For $\log \dsk$, however, a uniform prior, $p(\log \dsk) \propto 1$, yields an improper posterior
distribution because the likelihood converges to a positive constant as $\log \dsk$ goes to
$-\infty$. Therefore, we specify a Cauchy distribution for $p(\log \dsk)$,
\begin{equation}
  \log \dsk \sim \text{Cauchy}(\text{center}=9, \ \text{scale}=5). \label{eq:prior_logds}
\end{equation}
which is a broad, fat-tailed distribution covering all conceivable values for the path length that
we expect based on all sets of \ion{Fe}{13} intensities, with an example set of intensities shown
in Table~\ref{tab:model}.

We assume the parameters are independent a priori so that the joint prior distribution is
\begin{equation}
  \begin{split}
    p(\rvm, \paramk) &= p(\rvm) \ p(\paramk) \\
    &= p(\rvm) \ p(\log \Nek) \ p(\log \dsk).
  \end{split}
  \label{eq:prior}
\end{equation}
Here $\param$ is indexed by $k$, but $\rvm$ is not. This reflects the fact that, although $\paramk$
vary among the pixels, we expect the true emissivity (i.e., the true value of $\rvm$) to be an
underlying physical quantity that is the same for all pixels.

We consider two ways to fit the plasma parameters, $\Theta$, given the observed or simulated
intensities, $\Data$, while accounting for atomic uncertainty, $\mathcal{M}$. First we can analyze
each pixel separately in a sequence of pixel-by-pixel analyses. Although this may yield different
estimates of $\rvm$, the index of the preferred emissivity curve among the pixels, it allows us to
see if the intensities of each pixel give consistent information as to the best emissivity
curve(s). Alternatively, we can simultaneously analyze the intensities from all the pixels to
arrive at an overall estimate of the most likely emissivity curve. Using this strategy, uncertainty
can be quantified with a list of the most likely emissivity realizations from $\mathcal{M}$ (or
their indices, $\rvm$) along with their associated posterior probabilities.

We consider both the separate pixel-by-pixel and simultaneous analyses, and for each develop both
pragmatic and fully Bayesian approaches. Specifically, Section~\ref{sec:posterior} develops the
pragmatic and fully Bayesian approaches to the pixel-by-pixel analyses and
Section~\ref{sec:separateanalysis} describes the algorithms used to deploy these approaches.  The
simultaneous analysis and its algorithm are discussed in Section~\ref{sec:jointanalysis}.%}

\subsection{Pragmatic and fully Bayesian methods for separate pixel-by-pixel analysis} \label{sec:posterior}

Given the likelihood function in Eq~(\ref{eq:likelihood}) and the prior distribution in
Eq~(\ref{eq:prior}), the joint posterior distribution for $\rvm$ and $\paramk$ under the separate
pixel-by-pixel analyses is
\begin{equation}
  p(\rvm, \paramk | \datak) = \frac{L(\rvm, \paramk \mid \datak) \ p(\rvm, \paramk)}{p(\datak)},
  \label{eq:joint_posterior_sep}
\end{equation}
where $p(\datak) = \sum_{\rvm=1}^M \int L(\rvm, \paramk \mid \datak) \ p(\rvm, \paramk)
\ \mathrm{d}\paramk$.

Then the marginal posterior distribution $p(\paramk \mid \datak)$ can be obtained by summing over
$\rvm$,
\begin{equation}
  p(\paramk \mid \datak) = \sum_{\rvm=1}^M p(\rvm, \paramk \mid \datak).
\label{eq:posterior_marginal_theta}
\end{equation}
In this way, we are able to infer $\paramk$ accounting for uncertainties of the atomic data via the ensemble in $\mathcal{M}$.%}

\subsubsection{Pragmatic Bayesian method} \label{sec:pragmaticbayesian}

For the pragmatic Bayesian method, as described by \cite{lee_cal2011}, we assume that the observed
intensities are uninformative as to the most likely emissivities. That is, we do not take into
account the information in the intensities for narrowing the uncertainty in the choice of
emissivity realizations. Mathematically, this assumption can be written $p(\rvm \mid \datak) =
p(\rvm)$, i.e., $\rvm$ and $\datak$ are independent. Thus, the pragmatic Bayesian joint posterior
distribution of $\rvm$ and $\paramk$ is
\begin{align}
p(\rvm, \paramk \mid \datak) &= p(\paramk \mid \datak, \rvm) \ p(\rvm \mid \datak) \label{eq:joint_posterior_sep_fully} \\ 
                             &= p(\paramk \mid \datak, \rvm) \ p(\rvm),  \label{eq:joint_posterior_sep_prag}
\end{align}
and the marginal posterior distribution of $\paramk$ (from Eq~(\ref{eq:posterior_marginal_theta})) is
\begin{equation}
  \begin{split}
    p(\paramk \mid \datak) &= \sum_{\rvm=1}^M p(\rvm, \paramk \mid \datak) \\
    & = \sum_{\rvm=1}^M p(\paramk \mid \datak, \rvm) \ p(\rvm).
    \label{eq:posterior_marginal_theta_prag}
  \end{split}
\end{equation} 

The pragmatic Bayesian method accounts for atomic uncertainty in a conservative manner. The
assumption that $p(\rvm \mid \datak) = p(\rvm)$ ignores information in the intensities, $\datak$,
that may reduce uncertainty of atomic data represented by $\rvm$ and hence of $\paramk$. We now
consider methods that allow $\datak$ to be informative for $\rvm$.

\subsubsection{Fully Bayesian method} \label{sec:fullybayesian}

In contrast to the pragmatic Bayesian method, the fully Bayesian method, as described by
\cite{xu2014fully}, incorporates the potential information in the data (i.e., the intensities) to
learn about $\rvm$. The fully Bayesian joint posterior distribution of $\rvm$ and $\paramk$ is
given in Eq~(\ref{eq:joint_posterior_sep_fully}) and the marginal posterior distribution of
$\paramk$ is given by
\begin{equation}
  \begin{split}
    p(\paramk \mid \datak) &= \sum_{\rvm=1}^M p(\rvm, \paramk \mid \datak) \\
    & = \sum_{\rvm=1}^M p(\paramk \mid \datak, \rvm) \ p(\rvm \mid \datak)
    \label{eq:posterior_marginal_theta_fully}
  \end{split}
\end{equation}
where each $p(\paramk \mid \datak)$ is normalized so that $\sum_{\rvm=1}^M p(\rvm \mid \datak) = 1$.

Using Bayes' theorem, we can directly compute the probability of each emissivity realization,
$\rvm$, given the data in each pixel separately,
\begin{equation}
  p(\rvm \mid \datak) = \frac{ p(\datak \mid \rvm)}{\sum_{\rvm=1}^M p(\datak \mid \rvm)}.
  \label{eq:emissivity_posterior_separate}
\end{equation}
This is the marginal posterior probability among those emissivity realizations in
$\mathcal{M}$. Eq~(\ref{eq:emissivity_posterior_separate}) holds because each of the $m$ has the
same prior probability (see Eq~\eqref{eq:prior_m}).

The Bayesian posterior distribution in Eq~(\ref{eq:emissivity_posterior_separate}) allows the
observed intensities to be informative for the atomic physics, following the principles of Bayesian
analysis \citep{xu2014fully}. It enables us to use the intensities to determine which emissivity
realizations are more or less likely and averages over (posterior) uncertainty in emissivity
realizations.

\subsection{Algorithms for the separate pixel-by-pixel analyses} \label{sec:separateanalysis}

\subsubsection{Algorithms for pragmatic Bayesian in the separate pixel-by-pixel analyses} \label{sec:alg_for_prag_sep}

The Metropolis-Hastings (MH) algorithm \citep[e.g.,][]{hastings1970monte} is a general term for a
family of Markov chain simulation methods that are useful for sampling from Bayesian posterior
distributions. Let $p(\psi | X)$ be the target posterior distribution, using the notations in
Section~\ref{sec:Bayes}. A proposed $\psi^*$ is sampled from a proposal distribution $q(\psi^* |
\psi^{(t)})$ at iteration $t+1$. Calculating the acceptance probability, $\rho = \frac{p(\psi^* |
  X) \ q(\psi^{(t)} | \psi^*)}{p(\psi^{(t)} | X) \ q(\psi^* | \psi^{(t)})}$, we set $\psi^{(t+1)} =
\psi^*$ with probability $\min(\rho, 1)$ and set $\psi^{(t+1)} = \psi^{(t)}$ otherwise.

To obtain a Monte Carlo (MC) sample of $(\rvm, \paramk)$ from the pragmatic Bayesian posterior in
Eq~(\ref{eq:joint_posterior_sep_prag}), we first obtain a MC sample of the emissivity index,
$\{\rvm^{(1)}, \ldots, \rvm^{(L)}\}$, from its prior distribution, Eq~(\ref{eq:prior_m}). For each
$\rvm^{(\ell)}$, with $\ell=1,\dots,L$, we can then sample $\{\paramk^{[\ell, t]}, t = 1, \ldots,
T\}$ from $p(\paramk \mid \rvm^{(\ell)}, \datak)$ using the MH algorithm. This requires that we
specify the proposal distribution $q(\param^* | \param^{(t)})$. To do so, we first compute the
value of $\paramk$ that maximizes $\log p(\paramk \mid \rvm^{(\ell)}, \datak)$, i.e., the maximum a
posteriori (MAP) estimates, $\hat{\param}_k$, along with the $2\times2$ Hessian matrix evaluated at
the mode $\hat{\param}_k$, $H(\hat{\param}_k)$, for each $\rvm^{(\ell)}$. We then use $t_4 \left(
\paramk \mid \hat{\param}_k, (-H(\hat{\param}_k))^{-1} \right)$ as the MH proposal distribution,
where $t_\nu \left( x \mid \mu, \Sigma \right)$ is the density of a multivariate $t$ distribution
with $\nu$ degrees of freedom, mode $\mu$, and scale matrix $\Sigma$, evaluated at $x$. This type
of MH sampler is known as an independence sampler (\citealt{gilks1996introducing}). We run MH for
$T$ iterations, the last of which is taken as the MC sample corresponding to $\rvm^{(\ell)}$, i.e.,
$\paramk^{(\ell)} = \paramk^{[\ell, T]}$.

\subsubsection{Algorithms for fully Bayesian in the separate pixel-by-pixel analyses}

In the fully Bayesian separate pixel-by-pixel analyses, our aim is to obtain a MC sample from the
joint posterior distribution, Eq~(\ref{eq:joint_posterior_sep_fully}), and we propose three basic
strategies for doing this: (i) two-step MC with MH, described in Section~\ref{sec:2stepMCwithMH}
and Appendix~\ref{sec:2stepMCwithMH_appen}, (ii) two-step MC with a Gaussian approximation,
described in Appendix~\ref{sec:2stepMCwithG_appen}, and (iii) Hamiltonian MC (HMC), described in
Appendix~\ref{sec:HMC_appen}. Specifically, the first strategy uses the MH algorithm while the
second strategy makes a Gaussian approximation to the conditional distribution of $\paramk$ given
the sampled emissivity realization $\rvm$, respectively. Comparing the three strategies, the
two-step MC with MH is preferred because of the accuracy of estimates with moderate computation
time, while two-step MC with a Gaussian approximation may be faster (but less accurate) and HMC can
be more accurate (but slower) under certain conditions.

\subsubsection{Implementation of two-step MC with MH for fully Bayesian in the separate pixel-by-pixel analyses} \label{sec:2stepMCwithMH}

In order to implement the fully Bayesian method and to obtain a MC sample of $\paramk$ via
Eq~(\ref{eq:posterior_marginal_theta_fully}), we first evaluate
Eq~(\ref{eq:emissivity_posterior_separate}) for each $\rvm$ where
\begin{equation}
  p(\datak \mid \rvm) = \int  L(\rvm, \paramk \mid \datak) \, p(\paramk) \, \mathrm{d}\paramk
  \label{eq:data_given_emis}
\end{equation}
{is the Bayesian evidence conditional on a given emissivity. For each sampled $\paramk$, we
need only evaluate the likelihood for $\rvm = 1, \ldots, M$, and then renormalize the $M$
likelihood values by this weighted sum, which can be achieved via a two-step sampling as described
in this section.

The two dimensional integral in Eq~(\ref{eq:data_given_emis}) can be evaluated numerically using
the grid generated from the Trapezoidal Quadrature Rule (TQR), which is suitable for finite domain
quadrature \footnote{Package 'mvQuad' provides a collection of methods for (potentially)
  multivariate quadrature in R, and is available at
  \url{https://cran.r-project.org/web/packages/mvQuad/}.}. The Product-Rule is also used in the
construction of multivariate grids, which leads to an evenly designed grid.

The two dimensional quadrature can then be expressed as
\begin{multline}
  \int  L(\rvm, \paramk \mid \datak) \, p(\paramk) \, \mathrm{d}\paramk \\ = \sum_{i,j} w_{i,j} \,
  L(\rvm, \log \Nek^{(i)}, \log \dsk^{(j)} \mid \datak) \, p(\log \Nek^{(i)}, \log \dsk^{(j)})
\end{multline}
where nodes $(\log \Nek^{(i)}, \log \dsk^{(j)})$ and weights $(w_{i,j})$ are defined by the chosen
quadrature rule \footnote{TQR and Product-Rule are used in the construction of multivariate grids,
  where $\text{level}=5$ is a subcommand in the grid creating commander, which represents accuracy
  level, typically number of evaluation points for the parameters in each dimension.}. The integral
range of the two parameters is $(\hat{\param}_k - 3 \times \text{sdev}_k, \hat{\param}_k + 3 \times
\text{sdev}_k)$ where $\text{sdev}_k$ is a vector of the square root of the diagonal elements in
variance-covariance matrix $(-H(\hat{\param}_k))^{-1}$.

Having evaluated Eq~(\ref{eq:emissivity_posterior_separate}) at each $\rvm$, we can obtain a MC
sample of the emissivity index, $\{\rvm^{(1)}, \ldots, \rvm^{(L)}\}$. For each $\rvm^{(\ell)}$ we
sample from $p(\paramk \mid \datak, \rvm^{(\ell)})$ using an independence sampler exactly as
described in Section~\ref{sec:alg_for_prag_sep}. For each $\rvm^{(\ell)}$, we run the independence
sampler for $T$ iterations to obtain the MC sample corresponding to $\rvm^{(\ell)}$,
$\paramk^{(\ell)} = \paramk^{[\ell, T]}$. The detailed two-step MC with MH ($\mathcal{S}_{MH}$) is
given in Appendix~\ref{sec:2stepMCwithMH_appen}.

\begin{figure*}[t!]
  \centerline{\includegraphics[clip,width=0.85\linewidth,bb=134 70 686 522]{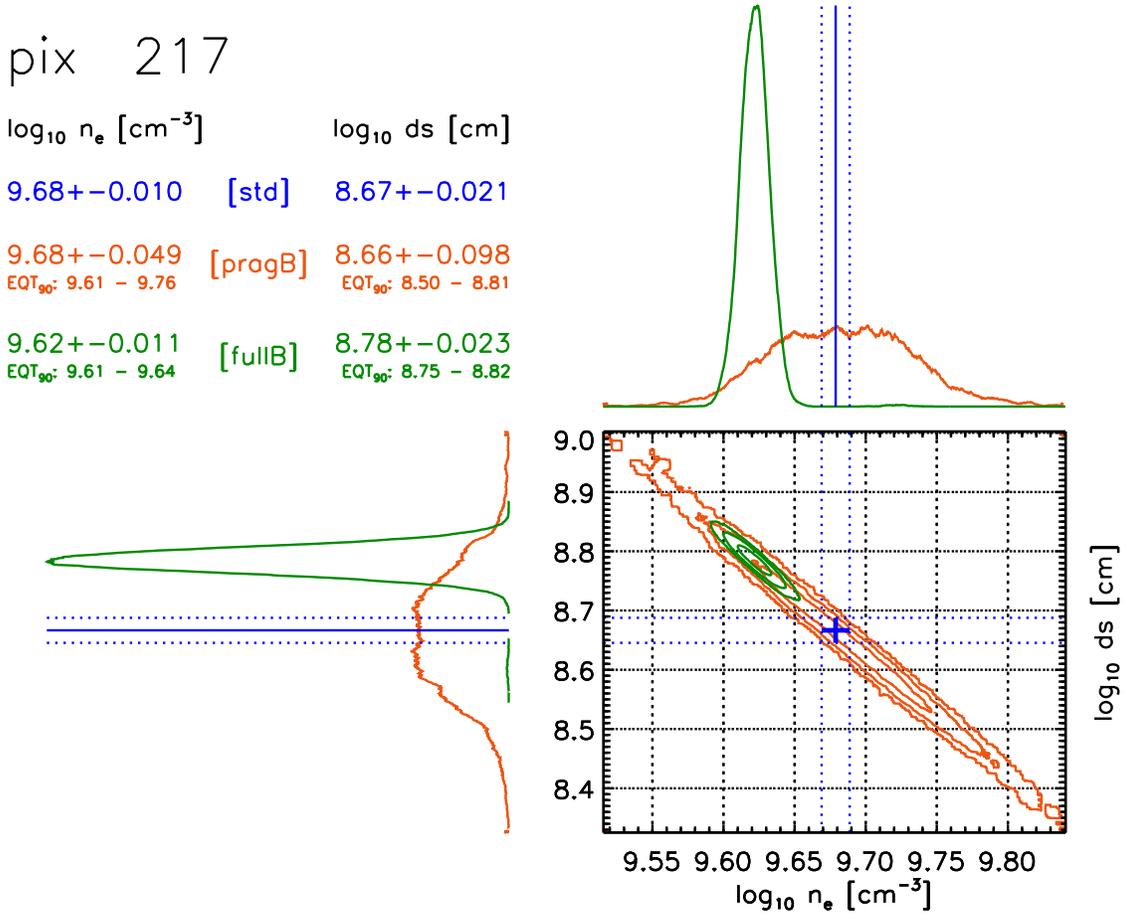}}
  \caption{{Comparisons of the inferred density $n_e$ and path length $ds$ using different methods
      for pixel \#$217$ (for the intensities listed in Table~\ref{tab:model}).  The results from
      the different methods used are color coded, with blue representing the standard method, red
      the pragmatic Bayes method (Section~\ref{sec:pragmaticbayesian}), and green the fully Bayes
      method (Section~\ref{sec:fullybayesian}) with the latter two computed for all pixels
        simultaneously (see Section~\ref{sec:jointanalysis}). The contour plots (with levels at
      $0.01\times$, $0.1\times$, and $0.5\times$ the maximum) show where the majority of the mass
      of the joint probability distributions of $(n_e,ds)$ fall, and their marginalized
      distributions along each axis is shown to the top and to the right of the corresponding
      axis. The results from the standard analysis is shown along with the histograms as straight
      lines (solid for the best-fit and dashed denoting the $\pm1\sigma$ errors on the best-fit
      obtained from the default CHIANTI emissivity functions), extending into the contour plot
      region.  The best-fit value from standard analysis is also marked on the contour plot with a
      '+' sign, with the arms of the symbol corresponding to the sizes of the error bars. The
      standard deviations of the marginalized posterior densities, as well as the $90$\% equal-tail
      bounds for both the pragmatic and full Bayes cases are listed in the legend.  As expected,
      the density and path length are highly correlated.  The standard method underestimates the
      uncertainties, the pragmatic Bayes method inflates them due to atomic data uncertainties.
      The fully Bayes method strikes a balance between atomic data uncertainties and how well the
      data are fit, shrinking the error bars relative to pragmatic Bayes and shifting the
      estimates.  The full set of plots for all $1000$ pixels considered here are available as a
      supplementary figure in the online journal.}}
  \label{fig:test1}
\end{figure*}

\subsection{Simultaneous analysis} \label{sec:jointanalysis}

When we consider all the $K$-pixel intensities together in a simultaneous analysis using the fully
Bayesian method, the likelihood function of $\rvm$ and $\paramj$ given $\Data$, and the prior
distribution of $\rvm$ and $\paramj$ are, respectively,
\begin{equation}
L(\rvm, \paramj \mid \Data) = \prod_{k=1}^K L(\rvm, \paramk \mid \datak)
\label{eq:likelihood_jointK}
\end{equation}
and
\begin{equation}
p(\rvm, \paramj) = p(\rvm) \ \prod_{k=1}^K p(\paramk).
\label{eq:prior_jointK}
\end{equation}
Thus, the joint posterior distribution of $\rvm$ and $\Theta$ can be expressed as 
\begin{equation}
  p(\rvm, \paramj \mid \Data) = \frac{L(\rvm, \paramj \mid \Data) p(\rvm, \paramj)}{p(\Data)},
  \label{eq:joint_posterior_jointK}
\end{equation}
where $p(\Data) = \sum_{\rvm=1}^M \int L(\rvm, \paramj \mid \Data) p(\rvm, \paramj) \ \mathrm{d}\paramj$. Similarly, treating $\rvm$ as an unknown parameter, we express the left hand side of Eq~(\ref{eq:joint_posterior_jointK}) as
\begin{equation}
  p(\rvm, \paramj \mid \Data) = p(\paramj \mid \Data, \rvm) \ p(\rvm \mid \Data),
  \label{eq:joint_posterior_jointK2}
\end{equation}} 
and we conduct statistical inference by obtaining a MC sample from this joint posterior
distribution.

First we can use all the data simultaneously to obtain the marginal posterior probability of each
emissivity realization $\rvm$,
\begin{equation}
  p(\rvm \mid \Data) = \frac{\prod_{k=1}^K p(\datak \mid \rvm)}{\sum_{\rvm=1}^M \prod_{k=1}^K
    p(\datak \mid \rvm)}.
  \label{eq:emissivity_posterior_joint}
\end{equation}
and sample $\rvm^{(\ell)}$, for $\ell=1, \ldots,L$, with weights given by the marginal posterior
probabilities in Eq (\ref{eq:emissivity_posterior_joint}) so that those favoured by the data are
sampled more frequently. The computation of $p(\datak \mid \rvm)$ for each $k$ and $\rvm$ is
discussed in Section~\ref{sec:2stepMCwithMH}.

For each sampled $\rvm$, we sample $\param$ from its
conditional posterior distribution
\begin{equation}
  \begin{split}
    p(\paramj \mid \Data, \rvm) &\propto L(\rvm, \paramj \mid \Data) \ p(\paramj)  \\ 
    &= \prod_{k=1}^K L(\rvm, \paramk \mid \datak) \ p(\paramk) \\ &= \prod_{k=1}^K \prod_{j=1}^J \calN
    \left(\left.I_{k \lambda_j} \ \right| \ \emisk \Nek^2 \dsk, \ \sigma^2_{k \lambda_j} \right)
    \\ & \times p(\log \Nek) \ p(\log \dsk).
  \end{split}
  \label{eq:condi_posterior_jointK}  
\end{equation}
as these $K$-pixel datasets were randomly selected from the observations indicated in
Section~\ref{sec:observations}, so that we can safely assume conditional independence among them.

Similarly, an MH sampler is used to obtain a correlated MC sample, $\{ \paramj^{[t]}, t = 1,
\ldots, T\}$, from $p(\paramj \mid \rvm^{(\ell)}, \Data)$. A $t_4 \left( \paramk \mid
\hat{\param}_k, (-H(\hat{\param}_k))^{-1} \right)$ proposal distribution is used for each pixel
independently and separately to make the computation more efficient. With this proposal
distribution, we run the MH for $T$ iterations over all the $K$-pixel intensities and obtain the MC
sampler corresponding to $\rvm^{(\ell)}$, $\paramj^{(\ell)} = \paramj^{[\ell, T]}$. The detailed
two-step MC with MH via simultaneous analysis ($\mathcal{S}_{MH_{simul}}$) is given in
Appendix~\ref{sec:joint_appen}.

\subsection{Application to simulated intensities} \label{sec:2stepMCwithMH_testset}

Here we illustrate both the separate pixel-by-pixel and the simultaneous analyses, mentioned in
Section~\ref{sec:separateanalysis} and Section~\ref{sec:jointanalysis}, with a simulated case,
using $K=1000$ simulated sets of intensities for each of $J=7$ spectral lines with known density
and path lengths as described in Section~\ref{sec:test}. This will allow for the comparison of our
inferred values with known values.

We run the separate pixel-by-pixel and simultaneous analyses described in
Section~\ref{sec:2stepMCwithMH} and Section~\ref{sec:jointanalysis}. For both analyses, $30$ MH
samplers, which is determined by constructing autocorrelation plots in this setting
\citep{xu2014fully}, are drawn for each sampled emissivity realization $\rvm^{(\ell)}$, and the
last MH sampler is taken as a MC sampler. There are $8000$ MC samplers drawn in each simulation.

\begin{figure*}[t!]
  %\begin{figure}[!htbp]
  \centerline{
    \includegraphics[width=0.5\linewidth]{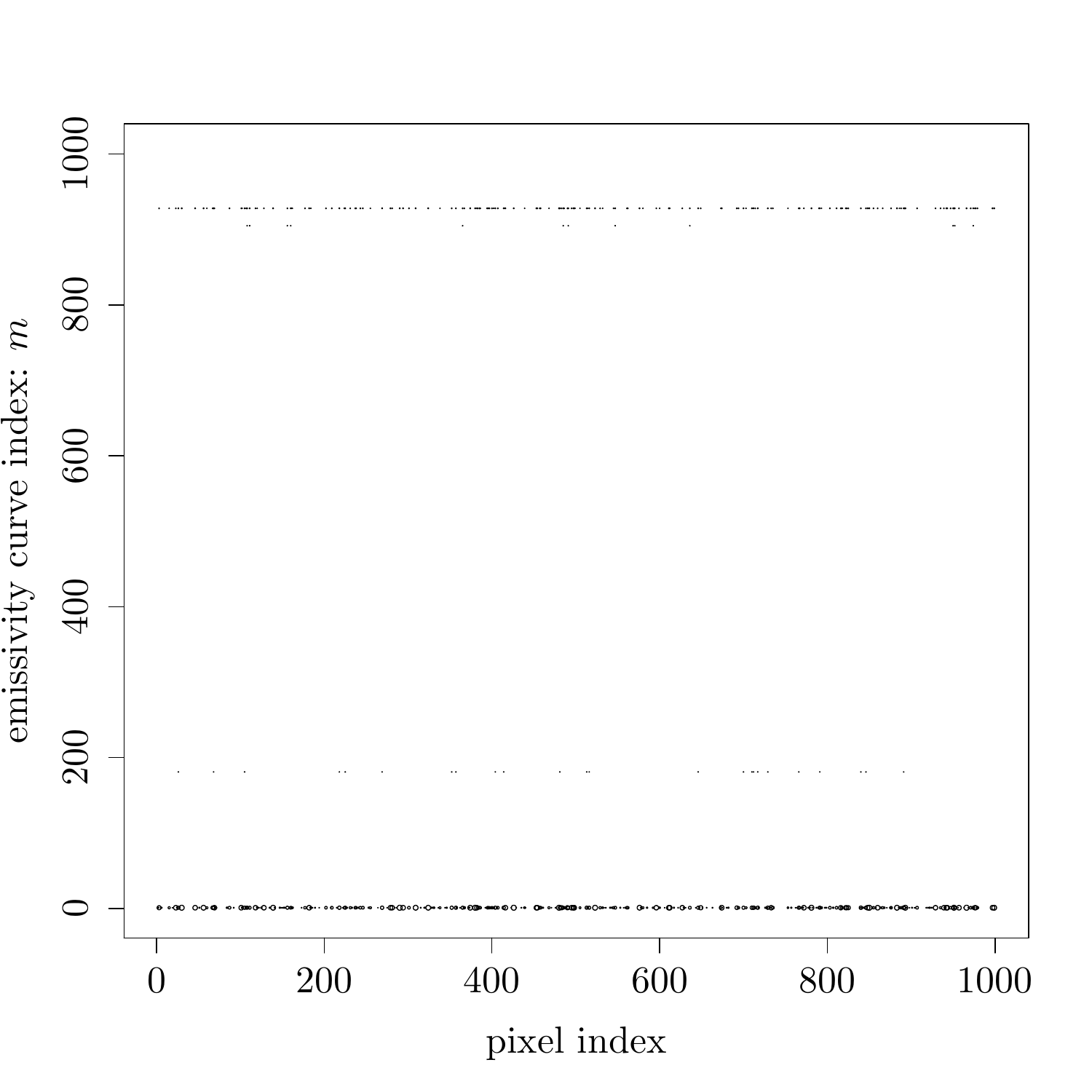}
    \includegraphics[width=0.5\linewidth]{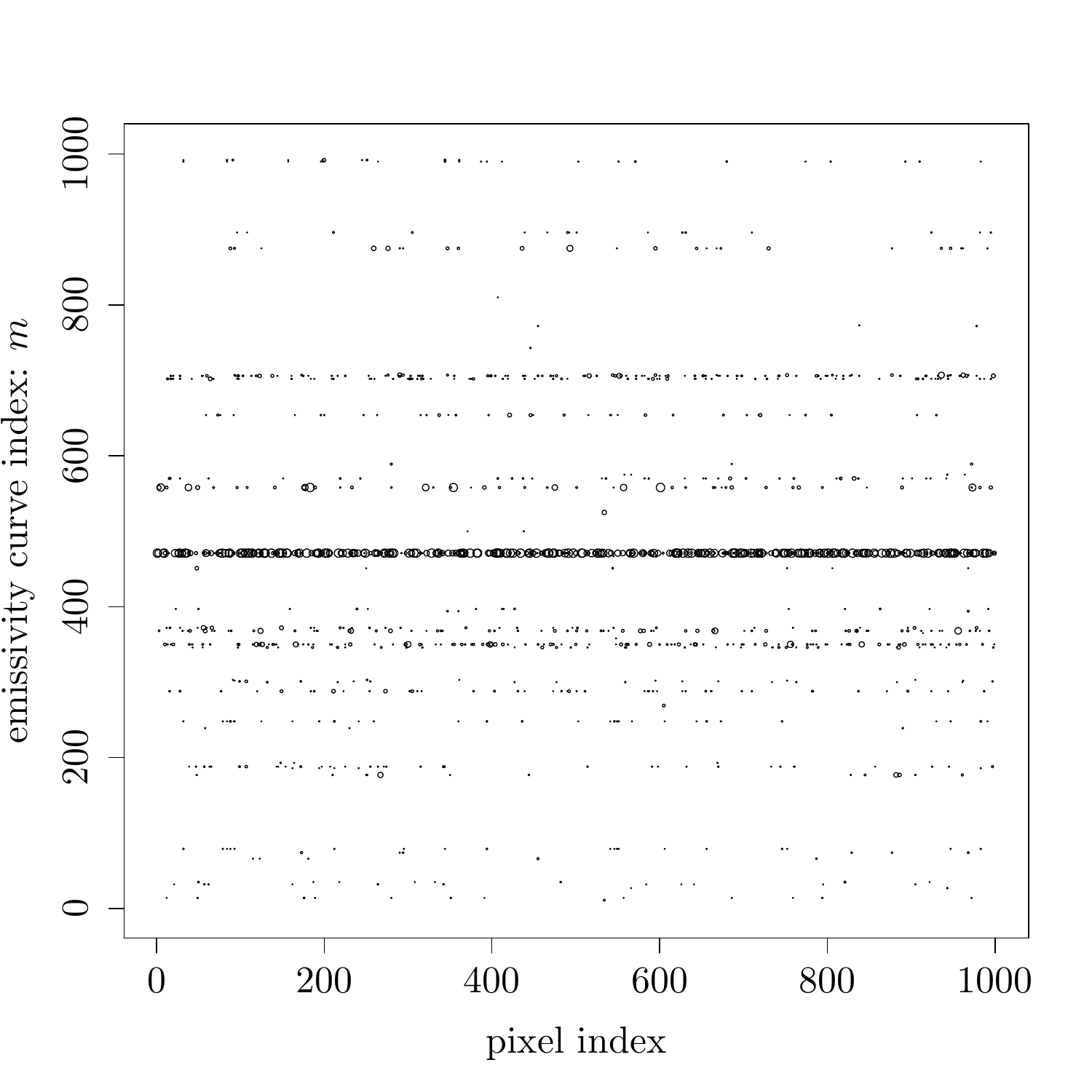}}
  \caption{Selecting the optimal emissivity curves with separate pixel-by-pixel fully
    Bayesian analysis. The x-axis and the y-axis represent the index of the pixels and the index of
    the emissivity curves respectively. For each pixel, the relative posterior probability is
    plotted along a vertical column for the emissivity indices, where index $1$ represents
    the default CHIANTI emissivities.  The size of the dots represents the relative values of the
    posterior probability ($p(\rvm \mid \datak)$, for emissivity index $\rvm$ and pixel dataset
    $\datak$) assigned to each emissivity index for a given pixel. The analyses carried out for the
    simulated dataset (left; generated using default $\rvm=1$, and showing only $p(\rvm \mid
    \datak)>0.06$) and for a real dataset (right; showing only $p(\rvm \mid \datak)>0.1$) are
    shown.}
  \label{fig:plot_m_given_data}
\end{figure*}

The comparison of the relative posterior probability $p(\rvm \mid \datak)$ for each emissivity
index and for each pixel, in separate pixel-by-pixel analyses, is shown in the left panel of
Figure~\ref{fig:plot_m_given_data}. The emissivity realization with index $1$ occupies almost all
of the probability. Similarly, in the simultaneous analysis, the posterior probability of the
emissivity realization with index $1$ is nearly to one. Both analyses recover the fact that all of
the simulated sets of intensities are computed from the actual CHIANTI atomic data (the emissivity
realization with index $1$) instead of the perturbed atomic data as described in
Section~\ref{sec:test}.

Comparing the results from separate pixel-by-pixel and simultaneous analyses using their mean
square errors (MSE), a measure of how well the fitted values explain the given set of observations,
Table~\ref{tab:mse_sep_joint} shows simultaneous analysis achieves smaller MSE values and indicates
the more data we have, the smaller MSE is achieved, i.e., simultaneous analysis gives a better
explanation of the given set of observations (i.e., intensities).

\begin{deluxetable}{lll}
  \tabletypesize{\scriptsize}
  \tablewidth{2.5in}
  \tablecaption{MSE between the fitted values and the true values for both parameters $(\log \Ne, \log \ds)$ via both separate pixel-by-pixel and simultaneous analyses \label{tab:mse_sep_joint}}
  \tablehead{
    \colhead{} &
    \colhead{$\log \Ne$} &
    \colhead{$\log \ds$}
  }
  \startdata
  $\mathcal{S}_{MH}$         & $1.345{\times}10^{-5}$ & $4.936{\times}10^{-5}$ \\
  $\mathcal{S}_{MH_{simul}}$ & $6.748{\times}10^{-7}$ & $2.241{\times}10^{-6}$ \\
  \enddata
\end{deluxetable}

\subsection{Application to observed intensities} \label{sec:2stepMCwithMH_observedset}

Here we demonstrate the effects of the different types of analyses by applying them to a real
dataset, the EIS full-CCD observations of an active region used as an exemplar in
Table~\ref{tab:model} (EIS file {\tt eis\_10\_20130708\_002042}). This dataset comprises sets of
measured intensities of $J=7$ spectral lines of $\mathrm{Fe \ XIII}$ in $K=1000$ distinct,
independent pixels. The results are shown in Figure~\ref{fig:test1} for the same pixel as
exemplified in Table~\ref{tab:test1}. The joint posterior probability density distribution
$p(\paramk \mid \datak)$ computed using the pragmatic and fully Bayesian methods are shown as
contour plots, and marginalized $1$-D posterior densities $p(\log \Nek \mid \datak)$ and $p(\log
\dsk \mid \datak)$ are shown as curves along the corresponding axes. The estimates of $\log \Nek$
and $\log \dsk$ computed via the standard analysis, i.e., the $\chi^2$ minimization of
Equation~(\ref{eq:model}), are marked with straight lines. Notice that the pragmatic Bayesian
method inflates the error bars relative to the standard method as it accounts for the atomic data
uncertainties. The fully Bayesian method shrinks the error bars relative to the pragmatic Bayesian
method and shifts the best estimate since it selects a subset of the full range of atomic
uncertainties that are consistent with the data. The standard method underestimates the
uncertainties in all cases, and is shifted relative to the fully Bayesian estimate.

The comparison of the relative posterior probability $p(\rvm \mid \datak)$ for each emissivity
index and for each pixel, in separate pixel-by-pixel analyses, is shown in the right panel of
Figure~\ref{fig:plot_m_given_data}. There are two dominant emissivity realizations which have a
combined posterior probability of over 0.99 using the two-step MC with MH. An example of the
posterior probability of the two dominant emissivity realizations given Pixel $593$ is shown in
Table~\ref{tab:posterioremis_singlepixel_2StepMCwithMH}. Similarly, in the simultaneous analysis,
the posterior probability of the emissivity curve with index $471$ is exactly one. It indicates
that the emissivity realizations reveal consistent feature of the solar atmosphere.

\begin{deluxetable}{lll}
  \tabletypesize{\scriptsize}
  \tablewidth{2.5in}
  \tablecaption{The posterior probability of the two dominant emissivity realizations given Pixel 593, $p(\rvm | \data_{593})$, using the two-step MC with MH via both separate pixel-by-pixel and simultaneous analyses.       
  \label{tab:posterioremis_singlepixel_2StepMCwithMH}}
  \tablehead{
    \colhead{\rvm} &
    \colhead{$\mathcal{S}_{MH}$} &
    \colhead{$\mathcal{S}_{MH_{simul}}$}
  }
  \startdata
  471    & $0.894$ & $1.000$ \\
  368    & $0.105$ & $0.000$ \\
  others & $<0.001$ & $0.000$ \\
  %others & $<0.0016$ & $0.000$ \\
  \enddata
\end{deluxetable}

The computational time is considered in terms of (i) the elapsed time and (ii) the sum of the user
and system times, which is a closer measure to real clock time.  For the separate pixel-by-pixel
analyses, the computation time over all 1000 pixels is about (i) 14.5 hours and (ii) 41.0 hours,
respectively, for the two measures of computational time. For the simultaneous analysis, both time
measurements are about 6.0 hours. These computation times consist of both the quadrature part and
sampling part; the computation of the quadrature part is exactly the same for both the separate
pixel-by-pixel and simultaneous analyses with a computation time of 1.2 hours for both
measurements.

\section{Conclusions and Discussion} \label{sec:discussion}

We have presented the first comprehensive treatment of atomic physics uncertainties in the analysis
of solar spectra. To make this analysis tractable, we have considered the relatively simple problem
of inferring the electron density and path length from a set of observed \ion{Fe}{13} intensities
and a simple model for the emission (see Equation~(\ref{eq:model})). For this work we have used
observed \ion{Fe}{13} intensities from the EIS spectrometer on the \textit{Hinode} satellite. If we
consider only the uncertainties due to counting statistics, we obtain very small error bars on the
electron density and path length, suggesting that the parameters are very precisely determined by
the observations.

An essential component of this analysis is a model that we have constructed for the uncertainties
in the collisional excitation and spontaneous decay rates. These rates are needed to compute the
plasma emissivities that relate the observed intensities with the physical parameters of the
plasma. This model for the uncertainties reflects the fact that for many transitions, such as those
between the lower levels in \ion{Fe}{13}, these rates appear to have converged. For other
transitions, however, the rates are still highly uncertain. We have modified the CHIANTI software
to produce self-consistent realizations of the atomic data based on this model for the
uncertainties.

We have used a Bayesian framework to interpret the observed intensities in the context of the
different realizations of the atomic data. A pragmatic Bayes approach, where each realization of in
the electron density and path length that are about a factor of 5 larger than the uncertainty
implied by counting statistics alone. A fully Bayesian approach, where we allow the observed
intensities to update the uncertainty in the emissivity curves, reduces the uncertainties in the
plasma parameters, but also suggests that a different realization of the atomic data is more likely
than the default CHIANTI calculation. This indicates some combination of systematic errors in the
atomic physics, instrument calibration, and the observed intensities.

The methodology that we have developed here is both labor intensive and computationally demanding.
Nevertheless, we believe that it represents a breakthrough in how atomic data uncertainties are
brought into an analysis.  Future improvements to the methodology and the structure of atomic
databases will no doubt improve the process and make it more accessible. The extension to other
emission lines would require an evaluation of the uncertainties in the collisional excitation and
spontaneous decay rates similar to those described in Section~\ref{sec:atodat} for each ion. Other
uncertainties, such as those for the ionization and recombination rates needed to compute the
ionization balances, also need to be addressed if lines from different ionization stages are
considered. Once these uncertainty models are determined, we can only generate discrete
realizations of the atomic data. This necessitates a brute force approach to computing the
posterior which includes a sum over all of the realizations. The more common scenario is that the
posterior is a continuous function of the parameters, which can be sampled more easily. It is
clear, however, that the uncertainty in the atomic data is often the dominant source of error in
the analysis of solar spectra. Thus this effort is essential to a rigorous analysis of the data.

Some constraints and the uncertainties in the atomic data could, in principle, be extracted from an
analysis of the probability distribution $p(m|D)$ of the different realizations. In practice,
however, our ability to consider this inverse problem is severely limited by the mismatch between
the very large number of rates that go into calculating the level populations: the modelled line
emissivities depend on 56394 rates and their associated uncertainties.  In principle, if all the
transitions produced by the main levels in the ion could be observed, some constraints could be
established. However, we only observe a very small number of emission lines.

Finally, we stress that the analysis presented here cannot overcome any limitations in the model
used to interpret the observations. In this work, for example, we have assumed that the observed
emission can be described by a simple model with a single density, temperature, and path
length. Despite its simplicity, this model reproduces the observed intensities remarkably well. The
path lengths, however, are relatively long ($ds \sim 10^9$\,Mm) compared to the path lengths
expected for the moss (see Equation~\ref{eq:ds}). It is likely that the observed emission is a
combination of high density, short path length emission from the moss and low density, long path
length emission from the overlying corona. To keep the analysis simple we have avoided using a more
complex model. However, it would be necessary to consider more complex emission measure
distributions if we seek to interpret the plasma parameters derived from the observations.

\acknowledgments The authors acknowledge the generous support from the International Space Science
Institute for hosting discussions among the ``Improving the Analysis of Solar and Stellar
Observations'' international team.  The full team consisted of 
Harry Warren (NRL; PI),
Mark Weber (CfA; co-PI),
Inigo Arregui (Inst. Astr. de Canarias),
Frederic Auchere (Inst. Astr. Spatiale),
Connor Ballance (Queen's Univ),
Jessi Cisewski (Yale),
Giulio Del Zanna (Cambridge),
Veronique Delouille (Royal Obs. Belgium),
Adam Foster (CfA),
Chloe Guennou (Columbia),
Vinay Kashyap (CfA),
Fabio Reale (OAPA Palermo),
Randall Smith (CfA),
Nathan Stein (UPenn/Spotify), 
David Stenning (IPA/SAMSI/Imperial), and
David van Dyk (Imperial).
GDZ acknowledges support from STFC via the the
University of Cambridge DAMTP astrophysics grant.  The UK APAP network was funded during 2012--2015
by STFC (grant No. ST/J000892/1).  HPW was supported by NASA's \textit{Hinode} project.  XY, DS,
and DvD were supported by a Marie-Skodowska-Curie RISE (H2020-MSCA- RISE-2015-691164) Grant
provided by the European Com- mission.  MW was supported under the Hinode/XRT contract NNM07AB07C
from MSFC/NASA to SAO. VLK was supported by NASA contract NAS8-03060 to the {\it Chandra X-ray Center}, and acknowledges travel support from the Smithsonian Competitive Grants Program for
Science Fund 40488100HH0043. CHIANTI is a collaborative project involving George Mason University,
the University of Michigan (USA) and the University of Cambridge (UK). Hinode is a Japanese
mission developed and launched by ISAS/JAXA, with NAOJ as domestic partner and NASA and STFC (UK)
as international partners; it is operated by these agencies in co-operation with ESA and the NSC
(Norway).

\appendix
\section{Appendix A} \label{sec:2stepMCwithMH_appen}

\subsection{Separate analyses: two-step MC with MH}

For Pixel $k$, i.e. the $k$th set of intensities, the two-step MC with MH ($\mathcal{S}_{MH}$)
proceeds for $\ell = 1, \ldots, L$ with

\begin{enumerate}[Step 1:]
\item Sample $\rvm^{(\ell)} \sim p(\rvm \mid \datak)$ via Eq~(\ref{eq:emissivity_posterior_separate}).
\item For $t = 1, \ldots, T-1$,
  \begin{enumerate}[Step 2.1:]
  \item Sample $\paramk^{\text{[prop]}} \sim t_4 \left( \paramk \mid \hat{\param}_k, (-H(\hat{\param}_k))^{-1} \right)$ and compute
    \begin{equation}
      \rho = \frac{p(\paramk^{\text{[prop]}} \mid \datak, \rvm^{(\ell)}) \ t_4 \left( \paramk^{[t]} \mid \hat{\param}_k, (-H(\hat{\param}_k))^{-1} \right) }{p(\paramk^{[t]} \mid \datak, \rvm^{(\ell)}) \ t_4 \left( \paramk^{\text{[prop]}} \mid \hat{\param}_k, (-H(\hat{\param}_k))^{-1} \right)}.
    \end{equation}
  \item Set
    \begin{equation}
      \paramk^{[t+1]} =
      \begin{cases}
        \paramk^{\text{[prop]}},& \text{with probability } \min(\rho, 1)\\
        \paramk^{[t]},              & \text{otherwise}
      \end{cases}.
    \end{equation}
  \end{enumerate}
\item Set $\paramk^{(\ell)} = \paramk^{[T]}$.
\end{enumerate}

For simplicity at each iteration, if the sampled emissivity index in Step 1 is the same as the
previous draw, we do not need to iterate MH to sample $\paramk$ in Step 2 since we already have
a good proposal distribution for the same target distribution. Moreover, if there does exist one
dominant emissivity curve, e.g., there exists $\rvm^*$ such that $p(\rvm^* \mid \datak) \ge
0.9999$, we only need to sample this $\rvm^*$ all the time.
\section{Appendix B} \label{sec:2stepMCwithG_appen}

\subsection{Separate analyses: two-step MC with Gaussian approximation} \label{sec:2stepMCwithG}

This is an alternative method to sample $p(\rvm, \paramk \mid \datak)$ based on
Eq~(\ref{eq:posterior_marginal_theta_fully}) and Eq~(\ref{eq:emissivity_posterior_separate}).  As
in Section~\ref{sec:2stepMCwithMH}, we can evaluate Eq~(\ref{eq:emissivity_posterior_separate}) at
each $\rvm$ and obtain a MC sample of the emissivity index, $\{\rvm^{(1)}, \ldots,
\rvm^{(L)}\}$. For each $\rvm^{(\ell)}$, instead of using exact MH algorithm, we can then sample
from $p(\paramk \mid \datak, \rvm^{(\ell)})$ by considering an approximate algorithm via Gaussian
approximation.

We can conduct a Gaussian approximation to $p(\paramk \mid \datak, \rvm^{(\ell)})$ with mean equal
to the MAP estimates, $\hat{\param}_k$, and variance-covariance matrix
$(-H(\hat{\param}_k))^{-1}$. Specifically the Gaussian approximation distribution $\calN
\left(\paramk \mid \hat{\param}_k, (-H(\hat{\param}_k))^{-1} \right)$ has the same mode and
curvature as the target conditional distribution $p(\paramk \mid \datak, \rvm^{(\ell)})$. Thus the
two-step MC with Gaussian approximation ($\mathcal{S}_{G}$) proceeds for $\ell = 1, \ldots, L$ with
\begin{enumerate}[Step 1:]
\item Sample $\rvm^{(\ell)} \sim p(\rvm \mid \datak)$ via Eq~(\ref{eq:emissivity_posterior_separate}).
\item Sample $\paramk^{(\ell)} \sim \calN \left(\paramk \mid \hat{\param}_k, (-H(\hat{\param}_k))^{-1} \right)$, where $\hat{\param}_k$ depends on $\rvm^{(\ell)}$.
\end{enumerate}

Similar to Section~\ref{sec:2stepMCwithMH_appen}, if there is one dominant emissivity curve, we
only need to sample this dominant one all the time.

\subsection{Results from the simulated set of intensities and the observed intensities}

Here we illustrate two-step MC with Gaussian approximation using a simulated case and a realistic
case as described in Section~\ref{sec:2stepMCwithMH_testset} and
Section~\ref{sec:2stepMCwithMH_observedset}.

For two-step MC with Gaussian approximation, same as two-step MC with MH in
Section~\ref{sec:2stepMCwithMH_testset}, TQR and Product Rule are used in computing multivariate
quadrature in Eq~(\ref{eq:data_given_emis}). Once we obtain a MC sample of emissivity index via
Eq~(\ref{eq:emissivity_posterior_separate}), a Gaussian approximation is conducted to $p(\paramk
\mid \datak, \rvm^{(\ell)})$ for each sampled $\rvm^{(\ell)}$ and each pixel $\datak$ as described
in Appendix~\ref{sec:2stepMCwithG}. There are $8000$ MC samplers drawn for each pixel.

The plots to compare the relative posterior probability for each emissivity index and for each
pixel are identical to those in the simulated and realistic cases in
Section~\ref{sec:2stepMCwithMH_testset} and Section~\ref{sec:2stepMCwithMH_observedset}.

In the realistic case, the computation time over all 1000 pixels is 8.0 hours or 20.7 hours, with
respect to the elapsed time or the sum of user and system times, respectively. It consists of both
the quadrature part and the sampling part, where the computation time of quadrature part is the
same as with the two-step MC with MH, 1.2 hours for both measurements.
\section{Appendix C} \label{sec:HMC_appen}

\subsection{Separate analyses: Hamiltonian Monte Carlo} \label{sec:HMC}

Another alternative method to obtain a MC sample from the joint posterior distribution in
Eq~(\ref{eq:joint_posterior_sep}) via the separate analyses is to start by obtaining a sample from
their marginal posterior distribution,
\[
\paramk^{(1)}, \ldots, \paramk^{(L)} \sim p(\paramk \mid \datak).
\]
First, we rewrite
\begin{equation}
  p(\paramk \mid \datak) \propto L(\paramk \mid \datak) \ p(\paramk),
  \label{eq:posterior_marginal_theta_2}
\end{equation}
where
\begin{equation}
  \begin{split}
    L(\paramk \mid \datak) &= \sum_{\rvm=1}^M L(\rvm, \paramk \mid \datak) \ p(\rvm \mid \paramk) \\
    &= \frac{1}{M} \sum_{\rvm=1}^M L(\rvm, \paramk \mid \datak) \\
    &= \frac{1}{M} \sum_{\rvm=1}^M \prod_{j=1}^J \calN \left(\left.I_{k \lambda_j} \ \right| \ \emisk \Nek^2 \dsk, \ \sigma^2_{k \lambda_j} \right),
  \end{split}
  \label{eq:likelihood_marginal}
\end{equation}
since the prior independent assumption, $p(\rvm \mid \paramk) = p(\rvm) = 1/M$, and the observation
independent assumption among lines of wavelengths.

Evaluating $p(\paramk \mid \datak)$ in this way we can use the Stan\footnote{Stan is a
  probabilistic modeling language developed by Andrew Gelman and collaborators. It interfaces with
  the most popular data analysis languages like R, Python, etc., and is available at
  \url{mc-stan.org}.} software package \citep{carpenter2016stan} to obtain $\{\paramk^{(1)},
\ldots, \paramk^{(L)}\}$ via HMC to sample directly from its marginal posterior distribution,
Eq~(\ref{eq:posterior_marginal_theta_2}). However, we must analytically marginalize over $\rvm$,
via Eq~(\ref{eq:likelihood_marginal}), since it cannot accommodate discrete parameters.

With these MC sample $\{\paramk^{(1)}, \ldots, \paramk^{(L)}\}$ in hand, we can sample $\rvm$ from
its conditional posterior distribution,
\begin{equation}
  \begin{split}
    p(\rvm \mid \paramk^{(\ell)}, \datak)
    &= \frac{p(\rvm) \ L(\rvm, \paramk^{(\ell)} \mid \datak)}{\sum_{\tilde{\rvm}=1}^M p(\tilde{\rvm}) \ L(\tilde{\rvm}, \paramk^{(\ell)} \mid \datak)} \\
    &= \frac{L(\rvm, \paramk^{(\ell)} \mid \datak)}{\sum_{\tilde{\rvm}=1}^M L(\tilde{\rvm}, \paramk^{(\ell)} \mid \datak)},
    \label{eq:posterior_m_condi_theta}
  \end{split}
\end{equation}
for $\ell = 1, \ldots, L$.

\subsection{Sampling multimodal posterior distributions with Stan} \label{sec:multimodalStan}

The simulation obtained in Appendix~\ref{sec:HMC} results in bimodal posterior distributions for
$\log \Ne$ and $\log \ds$ for a couple of pixel datasets. Specifically, the two modes correspond to
the two different emissivity curves. The resulting relative size of the two modes does not match
the actual posterior distributions indicating HMC algorithm has trouble in jumping between the
modes. This multiple-mode problem may be due to an insufficient number of emissivity curves because
our set of emissivities sample the full uncertainty range sparsely. To solve this problem, we have
experimented with adding a few strategically chosen synthetic emissivity curves to the set
$\mathcal{M}$ and the augmented set of curves is denoted by $\mathcal{M}^{\mathrm{aug}}$, where
$\mathcal{M}$ is a subset of $\mathcal{M}^{\mathrm{aug}}$, i.e., $\mathcal{M} \subset
\mathcal{M}^{\mathrm{aug}}$. These tend to connect the modes and allow HMC to jump between
modes. We can then remove the samples associated with the synthetic emissivity curves to get MC
samples purely from the original target.

We run the algorithm described in Appendix~\ref{sec:HMC} with $\mathcal{M}$ replaced by
$\mathcal{M}^{\mathrm{aug}}$. For each sampled value of $\paramk^{(\ell)}$, $\ell = 1, \ldots, L$,
we compute $p(\rvm \mid \paramk^{(\ell)}, \datak)$ for each $\rvm \in \mathcal{M}^{\mathrm{aug}}$,
with $\mathcal{M}$ replaced by $\mathcal{M}^{\mathrm{aug}}$ in
Eq~(\ref{eq:posterior_m_condi_theta}), and sample a value of $\rvm$, say $\rvm^{(\ell)}$, from
it. Once we have these sample values of $\rvm$, $\rvm^{(\ell)}$, for $\ell = 1, \ldots, L$, we can
then extract the samples of $\paramk$ that correspond to the non-synthetic emissivity curves to get
MC samples purely from the original target, i.e., consider the conditional posterior distribution
$p(\rvm \mid \paramk^{(\ell)}, \datak)$ for each $\rvm \in \mathcal{M}$.

This creative method of adding synthetic emissivity curves in HMC can be generalised to all pixel
datasets. If all the multiple-mode pixels have two modes and these two modes depend on the two same
emissivity curves, the same synthetic emissivity curves can be added into the original ones and the
above procedure can be repeated to all pixel datasets.

\subsection{Results from the simulated set of intensities and the observed intensities} \label{sec:multimodalStanResults}

Here we illustrate HMC with Stan through a simulated case and a realistic case as described in
Section~\ref{sec:2stepMCwithMH_testset} and Section~\ref{sec:2stepMCwithMH_observedset}.

For HMC with Stan ($\mathcal{H}$), a few strategically chosen synthetic emissivity curves are
added, as described in Appendix~\ref{sec:HMC} and Appendix~\ref{sec:multimodalStan}. There are $5$
chains running, $4000$ iterations each, and the first half of the iterations of each chain are
discarded as burn-in.

\begin{figure*}[t!]
  \centerline{
    \includegraphics[width=\linewidth]{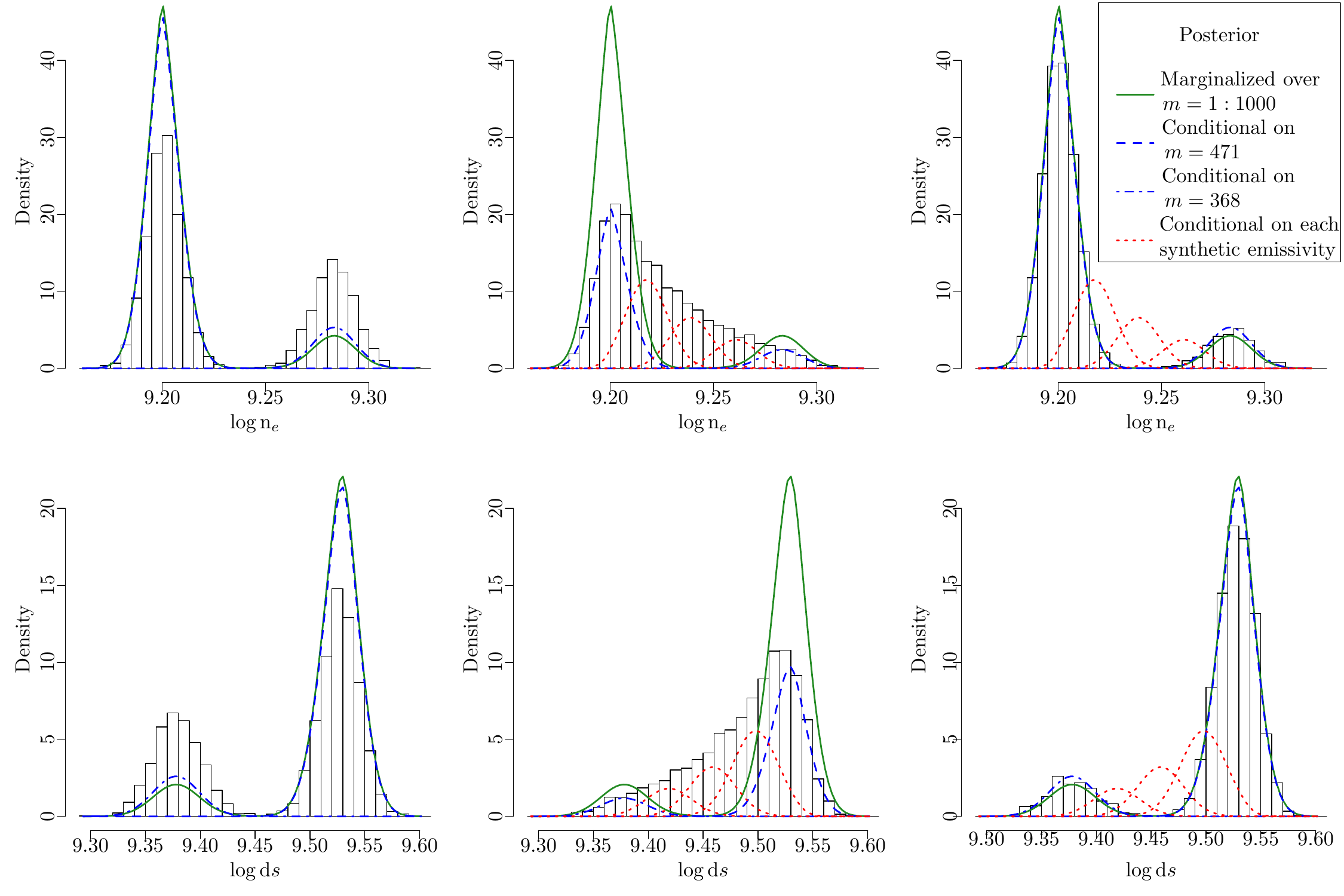}}
  \caption{ Demonstrating the augmented emissivity curves method to correct HMC/Stan analysis. The
    posterior probability densities are shown as calculated for the pixel $593$ dataset for the
    $\log\Ne$ (top row) and $\log\ds$ (bottom row). In all cases, the exact distributions,
    marginalized over the ensemble set of emissivity curves, are shown as as the solid green
    curves. The distributions conditional on one of the two most likely emissivity candidates,
    $\text{Emis}_{471}$ and $\text{Emis}_{368}$, are shown as the dashed and dot dashed blue lines
    respectively, and are renormalized with their corresponding posterior weights. The histogram
    represents the posterior density distributions computed using HMC via Stan. Going from left to
    right in each row, the left panels depict the problem that HMC produces modes of different
    heights compared to the exact calculation; the middle panels show its deformation as the
    augmented emissivity sample $\mathcal{M}^{\mathrm{aug}}$ is used where the posterior
    distributions conditional on each of the augmented emissivities are shown as the red dotted
    curves and are renormalized with their corresponding posterior weights; and the right panels
    show the corrected versions after removing the augmented emissivities.}
  \label{fig:Pixel593_syn}
\end{figure*}

In the simulated case, the comparison of the relative posterior probability $p(\rvm \mid \datak)$
for each emissivity index and for each pixel shows the emissivity curve with index $1$ occupies
almost all of the probability which also recovers the fact that all of the simulated sets of
intensities are computed from the actual CHIANTI atomic data (the emissivity curve with index $1$
instead of the perturbed atomic data as described in Section~\ref{sec:test}).

In the realistic case, once we run HMC with Stan as described in Appendix~\ref{sec:HMC}, bimodal
distributions appear for several of the pixels. The two modes correspond to two different
emissivity curves with index $471$ and $368$, i.e., $\text{Emis}_{471}$ and
$\text{Emis}_{368}$. Moreover, the relative size of the two modes does not match the actual
posterior distribution as shown in the left column of Figure~\ref{fig:Pixel593_syn}. Therefore, a
few strategically chosen synthetic emissivity curves are added to the original set and the
augmented set is
\[
\mathcal{M}^{\mathrm{aug}} / \mathcal{M} = \{w_1*\text{Emis}_{471} + w_2*\text{Emis}_{368}\}
\]
where $(w_1, w_2)=(0.75, 0.25)$, $(0.50, 0.50)$, and $(0.25, 0.75)$.  The HMC with Stan is run once
more with $\mathcal{M}$ replaced by $\mathcal{M}^{\mathrm{aug}}$ as described in
Appendix~\ref{sec:multimodalStan}. Samples of $\paramk^{(\ell)}$, $\ell = 1, \ldots, L$, are
obtained as shown in the middle column of Figure~\ref{fig:Pixel593_syn}. For each sampled value of
$\paramk^{(\ell)}$, we compute $p(\rvm \mid \paramk^{(\ell)}, \datak)$ for each $\rvm \in
\mathcal{M}^{\mathrm{aug}}$, via Eq~(\ref{eq:posterior_m_condi_theta}), and sample a corresponding
$\rvm^{(\ell)}$ from it. Considering the conditional posterior distribution $p(\rvm \mid
\paramk^{(\ell)}, \datak)$ for each $\rvm \in \mathcal{M}$, we can then extract the samples
$\paramk^{(\ell)}$ that correspond to the non-synthetic emissivity curves to get MC samples purely
from the original target as shown in the right column of Figure~\ref{fig:Pixel593_syn}. The
computation time over all $1000$ pixels is $51.4$ hours or $135.5$ hours, with respect to the two
ways of measuring the computation time, the elapsed time or the sum of user and system times
respectively.

\section{Appendix D} \label{sec:joint_appen}

\subsection{Simultaneous analysis}

The two-step MC with MH via simultaneous analysis ($\mathcal{S}_{MH_{simul}}$) proceeds for $\ell = 1,
\ldots, L$ with

\begin{enumerate}[Step 1:]
\item Sample $\rvm^{(\ell)} \sim p(\rvm \mid \Data)$ via Eq~(\ref{eq:emissivity_posterior_separate}).
\item Proceed for $t = 1, \ldots, T$,
  \begin{enumerate}[Step 2.1:]
  \item For each pixel $k=1, \ldots, K$, sample $\paramk^{\text{[prop]}} \sim t_4 \left( \paramk \mid \hat{\param}_k, (-H(\hat{\param}_k))^{-1} \right)$ \\
  and set $\paramj^{\text{[prop]}} = (\param_1^{\text{[prop]}}, \ldots, \param_K^{\text{[prop]}})$
  \item Compute 
    \begin{equation}
    \rho = \frac{\prod_{k=1}^K p(\paramk^{\text{[prop]}} \mid \datak, \rvm^{(\ell)}) \cdot \prod_{k=1}^K t_4 \left( \paramk^{[t]} \mid \hat{\param}_k, (-H(\hat{\param}_k))^{-1} \right) }{\prod_{k=1}^K p(\paramk^{[t]} \mid \datak, \rvm^{(\ell)}) \cdot \prod_{k=1}^K t_4 \left( \paramk^{\text{[prop]}} \mid \hat{\param}_k, (-H(\hat{\param}_k))^{-1} \right)}.
    \end{equation}
  \item Set
    \begin{equation}
      \paramj^{[t+1]} =
      \begin{cases}
        \paramj^{\text{[prop]}},& \text{with probability } \min(\rho, 1)\\
        \paramj^{[t]},              & \text{otherwise}
      \end{cases}.
    \end{equation}
  \end{enumerate}
\item Set $\paramj^{(\ell)} = \paramj^{[T]}$.
\end{enumerate}

Similar to the separate analyses, for simplicity at each iteration, if the sampled emissivity index
in Step 1 is not updated, we do not need to iterate MH to sample each $\paramk$ in Step 2. If there
does exist one dominant emissivity curve, we only need to sample the dominant all the time.

\begin{figure}[!htbp]
  \centerline{\includegraphics[width=5in,height=4.5in]{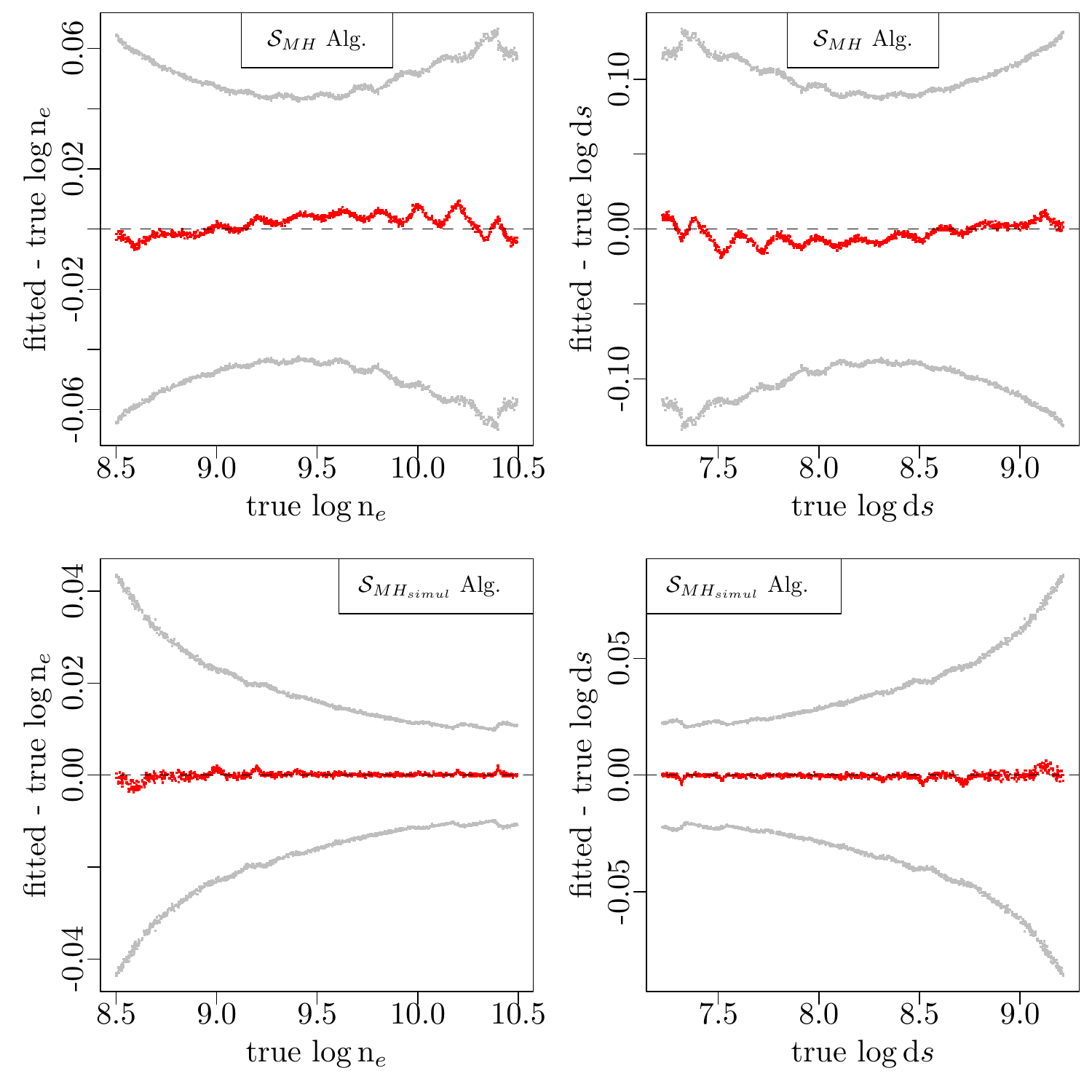}}
  \caption{ Comparison of best-fit values and actual input for the simulated dataset. The
    comparisons are shown for both $\log\Ne$ (left column) and $\log\ds$ (right
    column). Calculations are performed using fully Bayesian two-step MC with MH for each pixel
    dataset separately (top row) and for all the pixel datasets simultaneously (bottom row). The
    red dots represent the difference between the best-fit value and the actual input and the
    horizontal dashed lines represent the line of equality. The grey dots represent a vertical
    error of $\pm{1}$ standard deviation for the fitted values that incorporates atomic data
    uncertainty. Notice that the uncertainties are reduced when all the $K$-pixel datasets are used
    simultaneously.  }
  \label{fig:trueVSfitted}
\end{figure}

The results in Figure~\ref{fig:trueVSfitted} compare the fitted value using two-step MC with MH to
the true value of both parameters $\log \Ne$ (right) and $\log \ds$ (left) via both separate
pixel-by-pixel (top row) and simultaneous (bottom row) analyses. The grey lines represent the
vertical error of one standard deviation. The dashed line represents equality, where the fitted
value is identical to the true value. Compared with the separate pixel-by-pixel analyses, it shows
that the error bars are smaller around the truth when we use the simultaneous analysis than when we
use one pixel dataset at a time. The results in the plots illustrate that as more data are used in
the analysis by simultaneously analyzing those pixels, incorporating the uncertainty in the atomic
physics calculations results in more accurate fitted values.

\section{Appendix E}
\subsection{Comparison of Algorithms and Output Data Analysis} \label{appendix:comparisonofalgorithms}

To obtain a MC sample of the parameters, $\log \Nek$ and $\log \dsk$, via the separate
pixel-by-pixel analyses with joint posterior distribution in Eq~(\ref{eq:joint_posterior_sep}),
three algorithms were implemented for the fully Bayesian model on each of the $1000$ pixel observed
datasets: $\mathcal{S}_{G}$ in Appendix~\ref{sec:2stepMCwithG}, $\mathcal{S}_{MH}$ in
Section~\ref{sec:2stepMCwithMH}, and $\mathcal{H}$ in Appendix~\ref{sec:HMC}.

Our aim is to find which algorithm provides a more accurate simulation to the target posterior
distribution and is the best to be used to make statistical inference. From a statistical point of
view, we assume the HMC, which might give the best result, as the base line, and to see whether
these two two-step MC samplers provide better inference or not.

The first test statistic we consider is $z$-statistic, which is the difference in posterior mean
between the sample values from $\mathcal{S}_{G}$ or $\mathcal{S}_{MH}$ and from $\mathcal{H}$
divided by the standard deviation of $\mathcal{H}$ because HMC is assumed to be the base line,
indicating how far away that estimate is from the mean in standard units, i.e.,

\begin{equation}
  z_{\mathrm{score}}^i = \frac{\mathrm{mean}_{\mathcal{S}_i} -
    \mathrm{mean}_{\mathcal{H}}}{\mathrm{sd}_{\mathcal{H}}}, \ \text{for} \ i = G \text{ or } MH.
  \label{fn:zscore}
\end{equation}

\begin{figure}[!htbp]
  \centering
  \includegraphics[width=\linewidth]{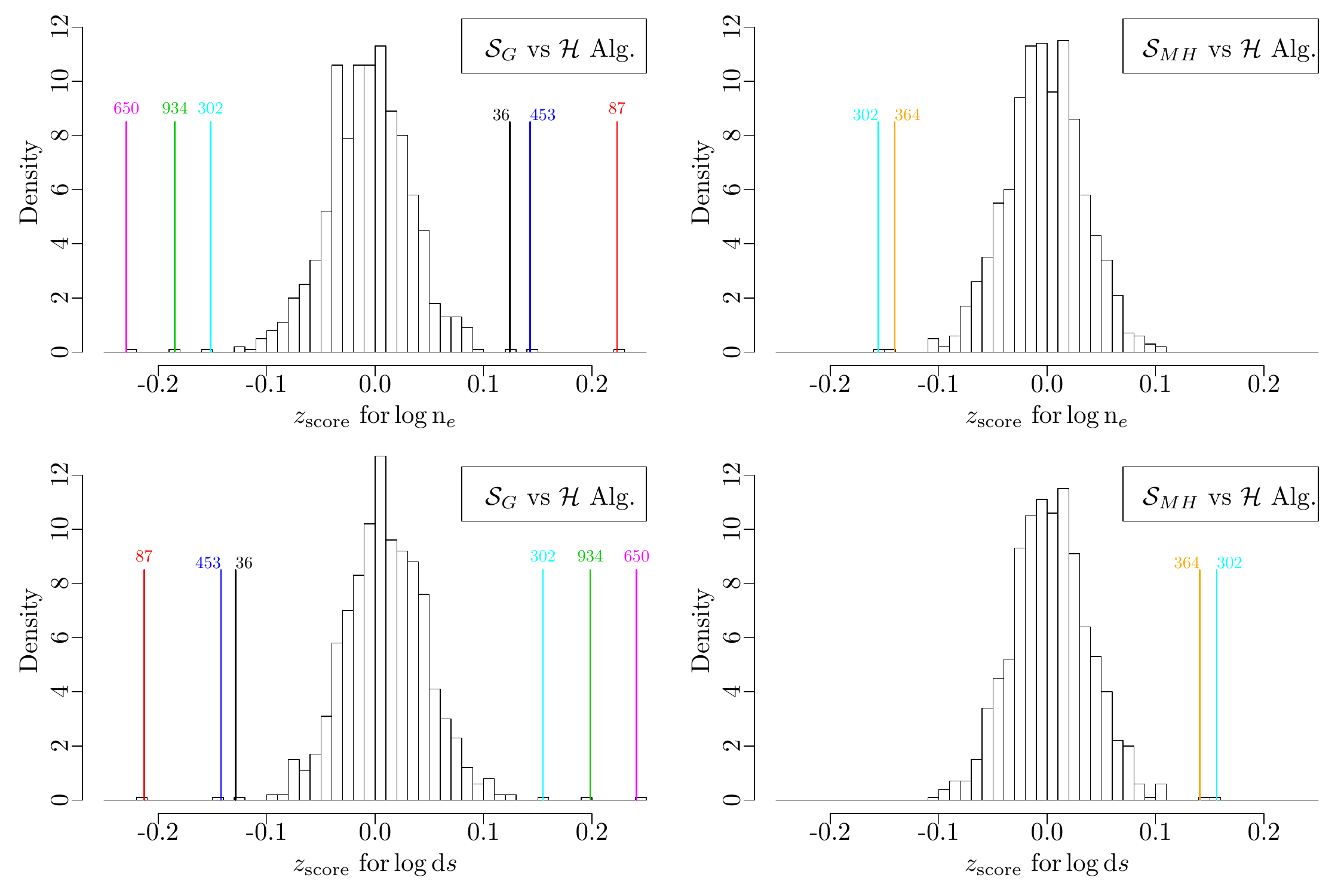}
  \caption{Selecting the extreme pixels via $z$-scores (unit difference in posterior mean) which
    are computed using the outputs of the three algorithms, $\mathcal{S}_{G}$, $\mathcal{S}_{MH}$,
    and $\mathcal{H}$ under fully Bayesian method and separate pixel-by-pixel analyses. The
    histograms represent the $z$-scores for both parameters, $\log \Ne$ (top row) and $\log \ds$
    (bottom row), in two comparisons (left: $\mathcal{S}_{G}$ to $\mathcal{H}$, right:
    $\mathcal{S}_{MH}$ to $\mathcal{H}$) respectively considering all the $1000$ pixels. The
    vertical lines correspond to the values of pixel indices, top left: $650, 934, 302, 36, 453,
    87$, top right: $302, 364$, bottom left: $87, 453, 36, 302, 934, 650$, bottom right: $364,
    302$, from left to right.}
  \label{fig:hist_Zscore}
\end{figure}

Figure~\ref{fig:hist_Zscore} shows the histograms of $z$-scores for both parameters, $\log \Ne$
(top row) and $\log \ds$ (bottom row), in two comparisons (left column: $\mathcal{S}_{G}$ to
$\mathcal{H}$, right column: $\mathcal{S}_{MH}$ to $\mathcal{H}$) respectively considering all the
$1000$ pixels. Looking at the worst case scenarios, the most extremes we see from the comparison on
the left-hand side is about $0.12$ to $0.25$ of standard deviation off, which corresponds to Pixel
$36$, $87$, $302$, $453$, $650$, and $934$. The comparison on the right-hand side indicates the
most extremes are about $0.15$ of standard deviation off occurring at Pixel $302$ and $364$. The
vertical lines correspond to the $z$-scores values of these extracted pixels. It suggests that we
need to look at the full posterior distributions for those extreme pixels and for the three
algorithms more closely, which will be found in Appendix~\ref{appendix:extracted_posteriorvalues},
to get some insights.

The second test statistic to compare is the ratio of standard deviations between $\mathcal{S}_{G}$
or $\mathcal{S}_{MH}$ and $\mathcal{H}$, i.e.,
\begin{equation}
  \frac{\mathrm{sd}_{\mathcal{S}_i}}{\mathrm{sd}_{\mathcal{H}}}, \ \text{for} \ i = G \text{ or } MH,
  \label{fn:ratioofsd}
\end{equation}
which essentially gives the relative size of confidence intervals that we compute.

\begin{figure}[!htbp]
  \centering
  \includegraphics[width=\linewidth]{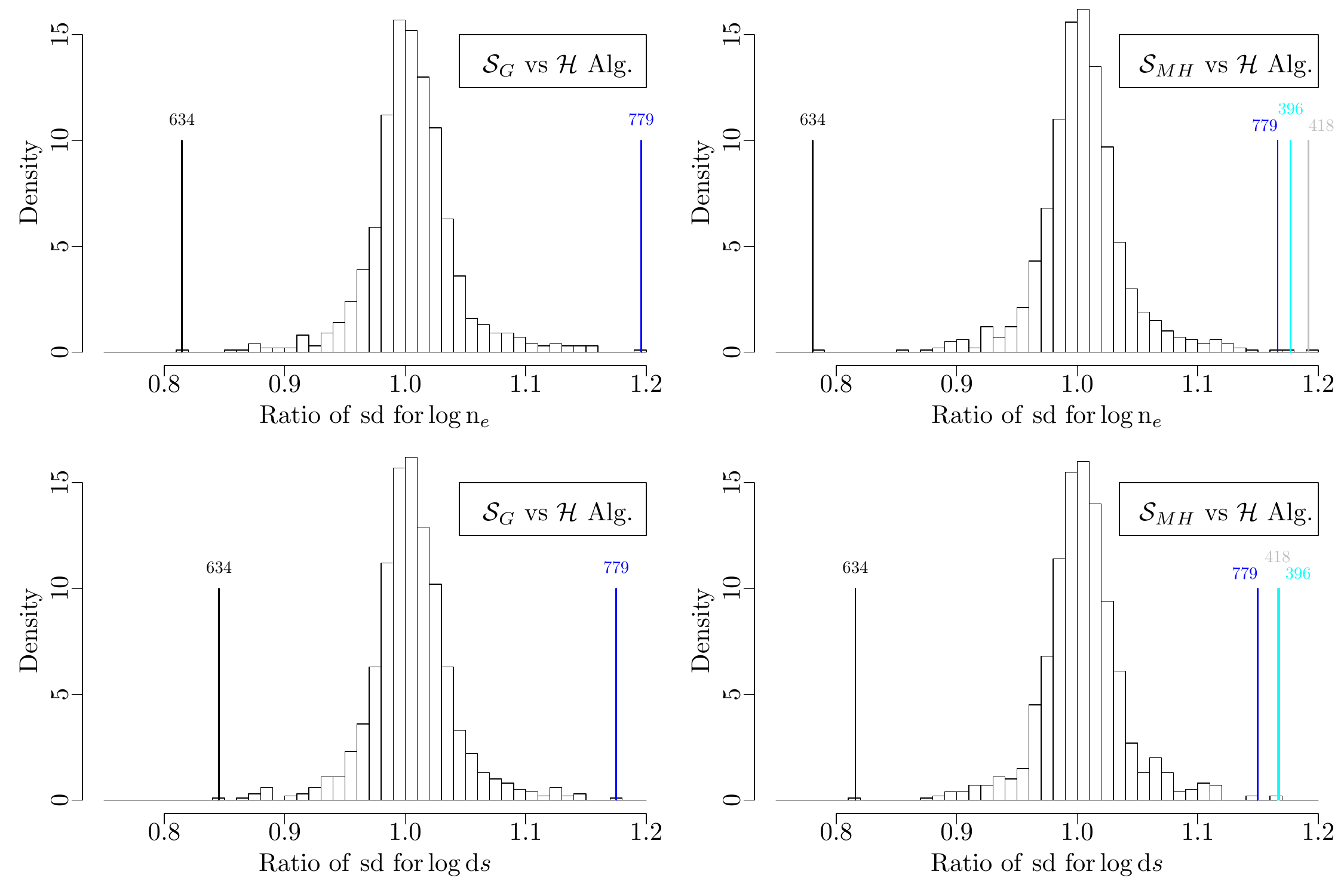}
  \caption{Selecting the extreme pixels via the ratio of standard deviations which are computed
    using the outputs of the three algorithms, $\mathcal{S}_{G}$, $\mathcal{S}_{MH}$, and
    $\mathcal{H}$ under fully Bayesian method and separate pixel-by-pixel analyses. The histograms
    represent the ratio of standard deviations for both parameters, $\log \Ne$ (top row) and $\log
    \ds$ (bottom row), in two comparisons (left: $\mathcal{S}_{G}$ to $\mathcal{H}$, right:
    $\mathcal{S}_{MH}$ to $\mathcal{H}$) respectively considering all the $1000$ pixels. The
    vertical lines correspond to the values of pixel indices, top left: $634, 779$, top right:
    $634, 779, 396, 418$, bottom left: $634, 779$, bottom right: $634, 779, 418, 396$, from left to
    right.}
  \label{fig:hist_sd}
\end{figure}

Figure~\ref{fig:hist_sd} shows the histograms of the ratio of standard deviations for both
parameters, $\log \Ne$ (top row) and $\log \ds$ (bottom row), in two comparisons (left column:
$\mathcal{S}_{G}$ to $\mathcal{H}$, right column: $\mathcal{S}_{MH}$ to $\mathcal{H}$) respectively
considering all the $1000$ pixels. The most extremes we see from the comparison on the left-hand
side corresponds to Pixel $634$ and $779$. The comparison on the right-hand side indicates the most
extremes occurring at Pixel $396$, $418$, $634$, and $779$. The vertical lines correspond to the
ratio values of these extracted pixels. An example of their posterior distributions for the three
algorithms can be found in Appendix~\ref{appendix:extracted_posteriorvalues}.

\subsection{Parallelization}

To improve the efficiency of the code, we parallelize the $1000$ pixels into $20$ or $10$
completely separate processes when pre-processing emissivities (i.e., obtaining the posterior
probability of each emissivity curve) or sampling $\param$, for all the three algorithms. The
$\mathrm{doParallel}$ package is used to provide a mechanism to execute $\mathrm{foreach}$ loops in
parallel within each process, where a multi-core backend is registered and a four worker cluster
(of a $64$-bit $2.5$ GHz CPU with $128$ GB of RAM) is created and used. Specifically, in the source
builds, we set the number of processors to use for the build to the number of cores on our machine
we want to devote to the build, which is thirty-two. We also set the maximum allowed number of
additional R processes allowed to be run in parallel to the current R processes, which is
thirty-two as well. For $\mathcal{H}$, each pixel is run with multiple cores and four pixels are
run at the same time. For $\mathcal{S}_{G}$ or $\mathcal{S}_{MH}$, we run each pixel with a
different core and thirty-two multi-core backends are used in parallel.

\subsection{The posterior values of the parameters for the three algorithms and for the extracted
  pixels} \label{appendix:extracted_posteriorvalues}

By comparing the three algorithms using the two test statistics mentioned in
Appendix~\ref{appendix:comparisonofalgorithms}, several extreme pixels are picked out from each
comparison.

\begin{sidewaysfigure}
  \centerline{
    \includegraphics[width=3in,height=6in]{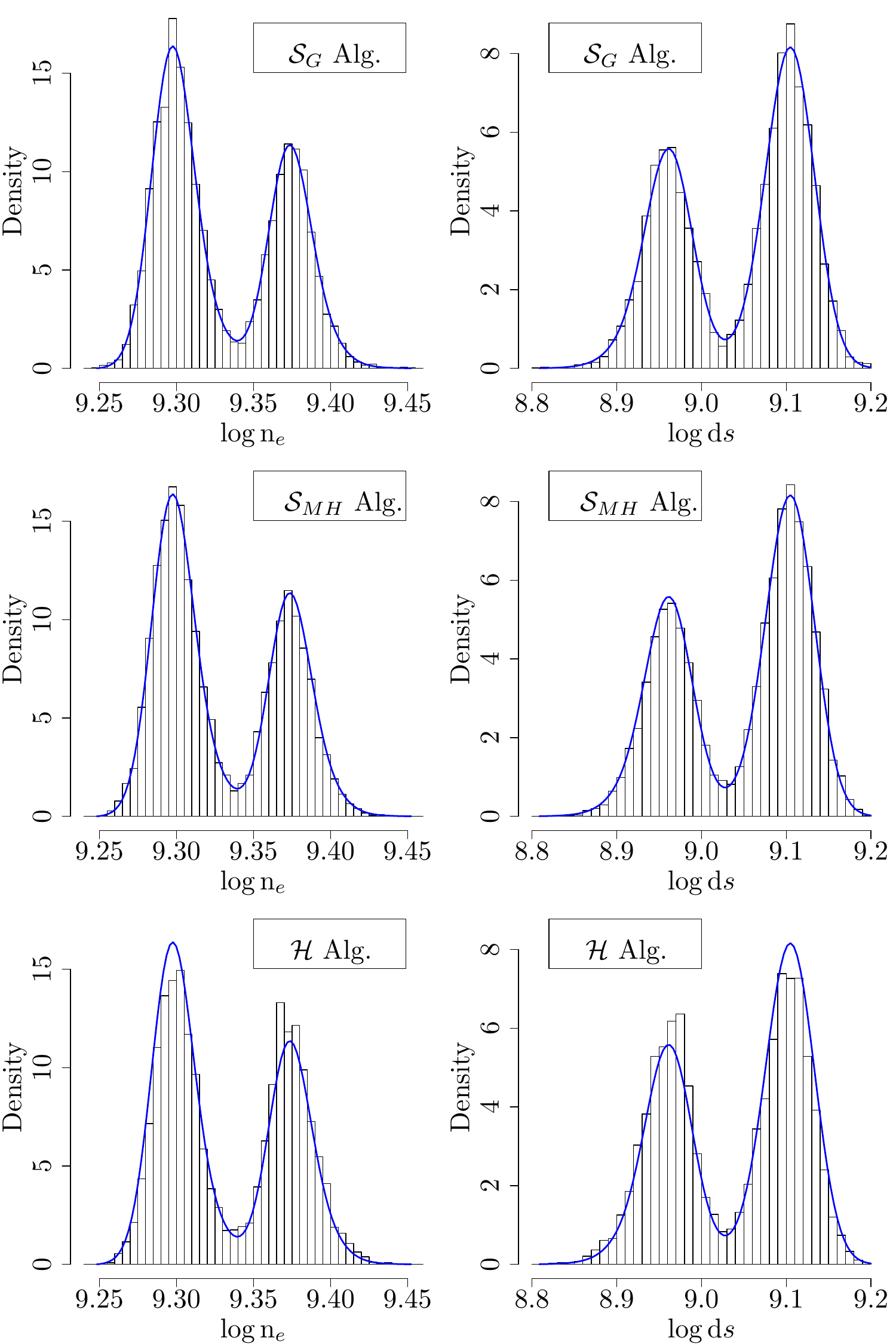}
    \unskip\ \vrule\
    \includegraphics[width=3in, height=6in]{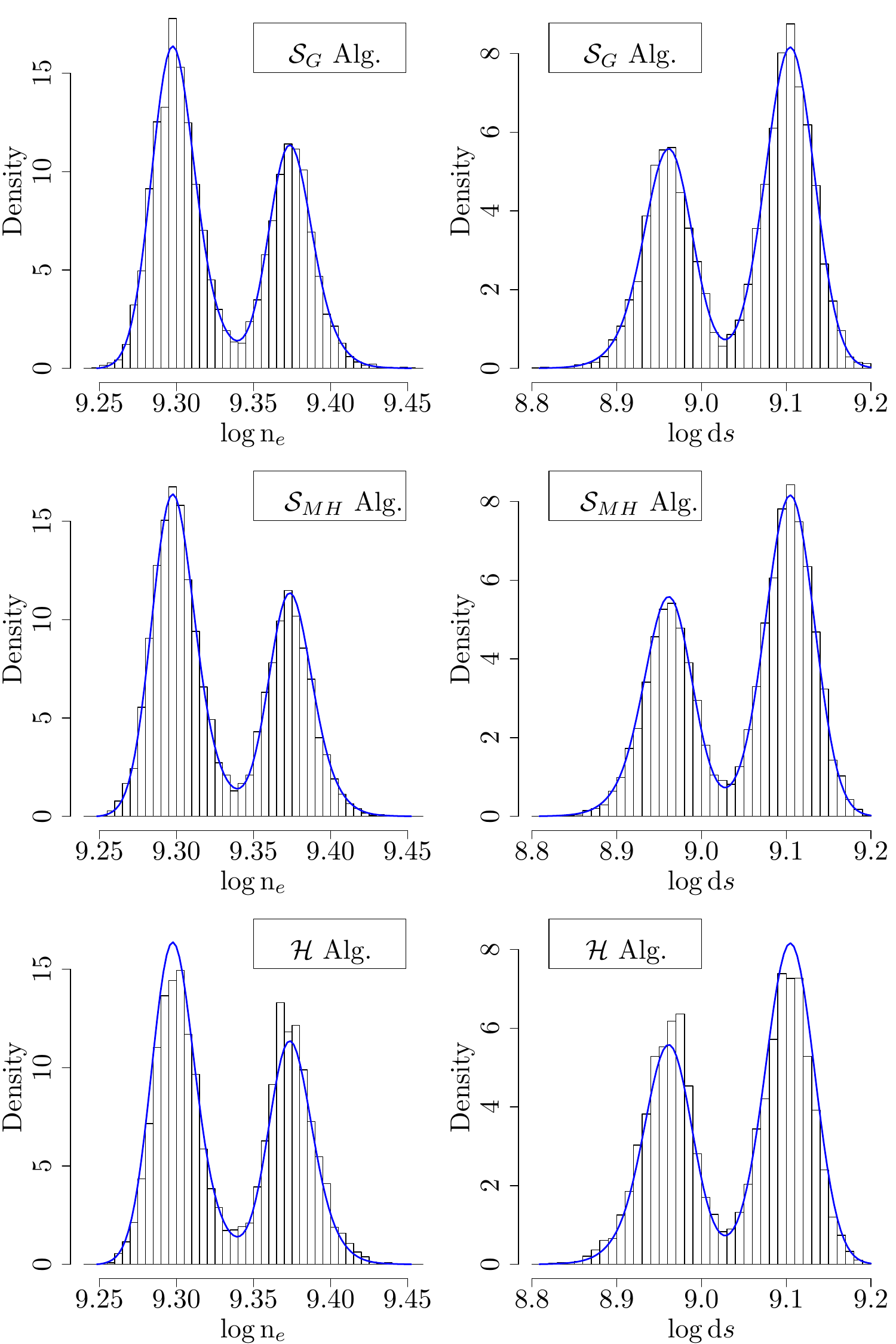}
    \unskip\ \vrule\
    \includegraphics[width=3in, height=6in]{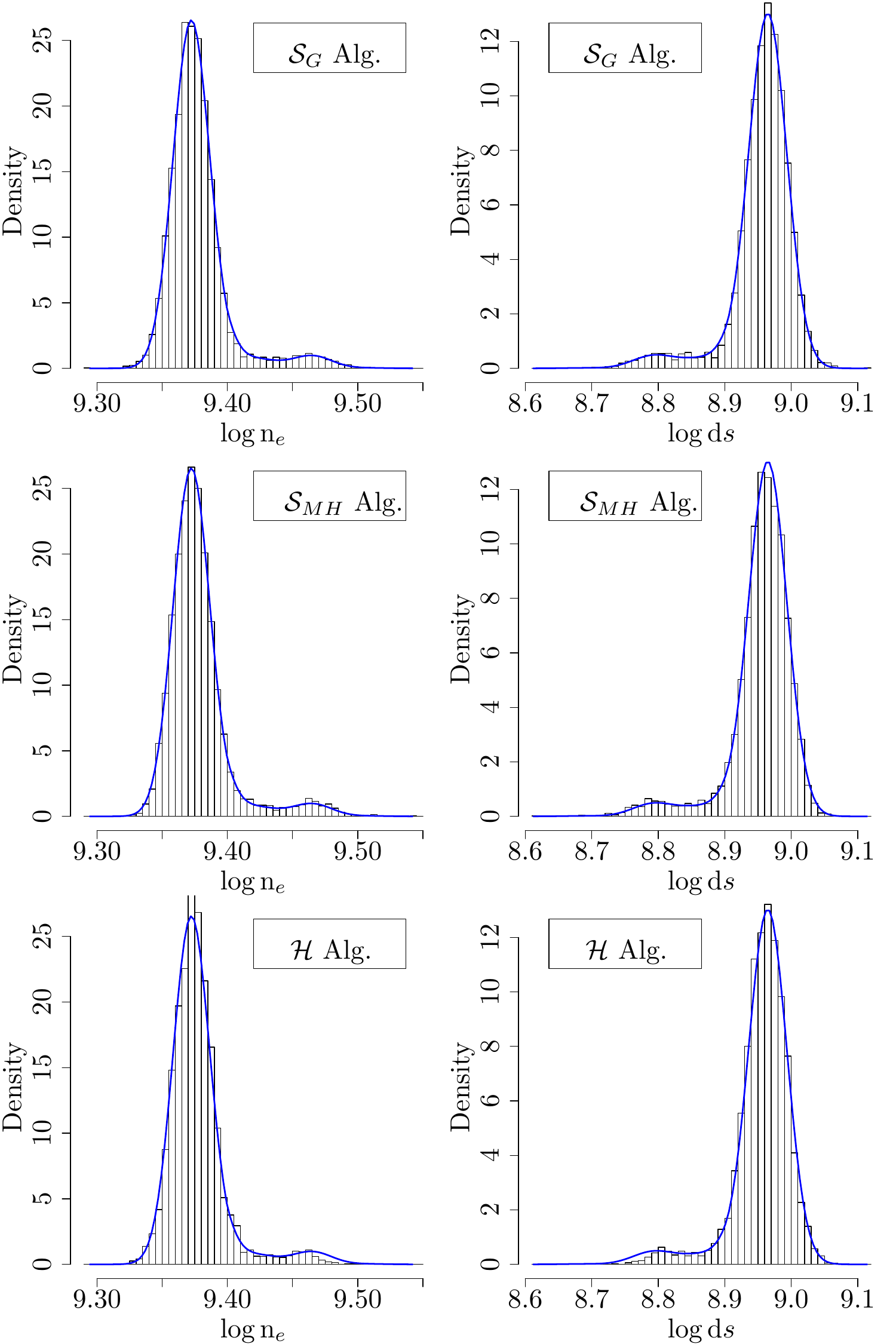}
  }
  \caption{Selecting the optimal algorithm from the full posterior distribution for the selected
    extreme pixels under fully Bayesian method. The histograms represent the posterior
    distributions of $\log \Ne$ (left column of each panel) and $\log \ds$ (right column of each
    panel) conditional on all $1000$ emissivity curves and Pixel $\mathbf{364}$ (left panel), or
    $\mathbf{396}$ (medium panel), or $\mathbf{650}$ (right panel) in real observed data set
    respectively. The sampling algorithms compared are $\mathcal{S}_{G}$ (first row),
    $\mathcal{S}_{MH}$ (second row), and $\mathcal{H}$ (third row). The actual conditional
    posterior distributions for each parameter are shown as the blue lines.}
  \label{fig:hist_pixels}
\end{sidewaysfigure}

Figure~\ref{fig:hist_pixels} show the histograms of the posterior values of the parameters $\log
\Ne$ (left) and $\log \ds$ (right) conditional on all $1000$ emissivity curves and the certain
extracted pixel dataset respectively. The sampling algorithms used are $\mathcal{S}_{G}$ algorithm
(top row), $\mathcal{S}_{MH}$ algorithm (middle row), and $\mathcal{H}$ algorithm (bottom
row). Three more synthetic emissivity curves are conditioned when using $\mathcal{H}$ as described
in Appendix~\ref{sec:multimodalStanResults}.

For Pixel $364$ (the left panel of Figure~\ref{fig:hist_pixels}), which are extracted from
the right column of Figure~\ref{fig:hist_Zscore}, having used the synthetic emissivity curves, it
is still not very great job of jumping between the modes for $\mathcal{H}$ algorithm in this
bimodal case.

For Pixel $396$ (the middle panel of Figure~\ref{fig:hist_pixels}), it is the histograms of
$\mathcal{H}$ algorithm that does not quite get into the tail that makes the standard deviation
from $\mathcal{H}$ algorithm relatively small and filters this pixel out from the right column of
Figure~\ref{fig:hist_sd}.

Similarly, for Pixel $650$ (the right panel of Figure~\ref{fig:hist_pixels}), which are extracted
from the left column of Figure~\ref{fig:hist_Zscore}, the $\mathcal{S}_{G}$ algorithm does not do a
great job at recovering the actual posterior with a noticeable discrepancy in the mode.

\subsection{Discussion}

As an example of the posterior distribution of emissivity curves given Pixel $593$, we get these
two dominant emissivity curves from both two-step MC samplers and HMC. Considering the two of them,
it is nearly all the probability up to $0.99$, as is shown in
Table~\ref{tab:posterioremis_singlepixel}.

\begin{deluxetable}{lcc}
  \tabletypesize{\scriptsize}
  \tablewidth{3.5in}
  \tablecaption{The posterior probability of the two dominant emissivity curves given Pixel $593$ using two-step MC and HMC via separate pixel-by-pixel analyses under fully Bayesian method \label{tab:posterioremis_singlepixel}}
  
  \tablehead{
    \colhead{\rvm} &
    \colhead{HMC with Stan ($\mathcal{H}$)} &
    \colhead{2stepMC with $\mathcal{S}_{G}$ or $\mathcal{S}_{MH}$}
  }
  \startdata
  471 & 0.860 & 0.894 \\
  368 & 0.138 & 0.105 \\
  others & $<0.0019$ & $<0.0016$
  \enddata
\end{deluxetable}

The computation time, in the realistic case, over all 1000 pixels for the three algorithms are
shown in Table~\ref{tab:computationtime}. Two ways of measuring the computation time are included,
the elapsed time and the sum of user and system times. Moreover, the computation time of both
two-step MC samplers consist of both quadrature part and sampling part, where the computation times
of quadrature part is $1.2$ hours for both measurements.

\begin{deluxetable}{lcc}
  \tabletypesize{\scriptsize}
  \tablewidth{3.5in}
  \tablecaption{Computation time for the three algorithms in realistic case under fully Bayesian method and separate pixel-by-pixel analyses \label{tab:computationtime}}
  \tablehead{
    \colhead{Algorithm} &
    \colhead{Elaplsed (h)} &
    \colhead{Sum of User$\&$System (h)}
  }
  \startdata
  $\mathcal{S}_{MH}$ & $14.5$ & $41.0$ \\
  $\mathcal{S}_{G}$  &  $8.0$ & $20.7$ \\
  $\mathcal{H}$      & $51.4$ & $135.5$ \\
  \enddata
\end{deluxetable}

We actually have two basic strategies for obtaining a MC sample from the joint posterior
distribution, HMC and two-step MC sampler. By comparing the histograms of the posterior values,
there is definitely an issue with the Gaussian assumption (two-step MC with Gaussian approximation)
where the MC samplers are not matching very well with the actual posterior and it is more
conservative. HMC algorithm looks appropriate but occasionally does not give the relative size of
the mode right, though after adding synthetic emissivity curves. For all 1000 pixels, the MC
samplers generated from two-step MC with MH match the density line of actual posterior very well
and this algorithm takes moderate computation time.  Therefore, more accurate and significantly
faster, two-step MC with MH would be the best to use to make statistical inference.

In our experiment, there are $J=7$ spectral lines with corresponding wavelengths being considered,
whereas two of them are not close to others in wavelength, $196.525$ and $209.916$ vs
$200.021$-$203.826 \ \mathring{\text{A}}$, and we call them extreme wavelengths. We have
experimented with one of the two-mode-case pixels (Pixel 593), where the two extreme wavelengths
are removed one at a time from the analysis and the three algorithms mentioned in
Section~\ref{sec:2stepMCwithMH}, Appendix~\ref{sec:2stepMCwithG}, and Appendix~\ref{sec:HMC} are
repeated.  Whether we consider the two extreme wavelengths or not, the resulting MC samplers have a
good match to their actual posterior distributions; however, the shape of the actual posterior
distribution differs dramatically when including wavelength $196.525 \mathring{\text{A}}$ compared
to when it is excluded from the analysis. Because including the extreme wavelengths did not impact
the ability of the MC samplers to recover the actual posterior distributions, we used the
seven-wavelength dataset in all the experiments.

%% ------------------------------------------------------------------------------------------
%% --- REFERENCES ---------------------------------------------------------------------------
%% ------------------------------------------------------------------------------------------

\end{document}